\documentstyle[12pt]{report}
\begin{document}

\newcommand{\be}{\begin{equation}}
\newcommand{\ee}{\end{equation}}
\newcommand{\bea}{\begin{eqnarray}}
\newcommand{\eea}{\end{eqnarray}}
\newcommand{\lb}{\label}
\newcommand{\hs}{\hspace}
\textwidth=14.5cm
\textheight=22cm
\intextsep=1.5cm
\oddsidemargin=0.5cm
\topmargin=-1cm
\setlength{\unitlength}{1mm}

\begin{titlepage}
\begin{flushright}
Freiburg THEP-93/27 \\
gr-qc/9312015
\end{flushright}
\begin{center}

{\large\bf THE SEMICLASSICAL APPROXIMATION TO
QUANTUM GRAVITY}\footnote{To appear in {\em Canonical gravity --
from classical to quantum}, edited by J. Ehlers and H. Friedrich
(Springer, Berlin, 1994).}

\vskip 1cm
{\bf Claus Kiefer}
\vskip 0.5cm
Fakult\"at f\"ur Physik, Universit\"at Freiburg,
Hermann-Herder-Str. 3,\\D-79104 Freiburg, Germany
\end{center}
\vskip 2cm
\begin{center}
{\bf Abstract}
\end{center}
\begin{quote}
A detailed review is given of the semiclassical approximation
to quantum gravity in the canonical framework. This includes in
particular the derivation of the functional Schr\"odinger equation
and a discussion of semiclassical time as well as the derivation
of quantum gravitational correction terms to the Schr\"odinger
equation. These terms are used to calculate energy shifts
for fields in De~Sitter space and non-unitary contributions
in black hole evaporation. Emphasis is also put on the relevance
of decoherence and correlations in semiclassical gravity.
The back reaction of non-gravitational quantum fields onto
the semiclassical background and the emergence of a Berry connection
on superspace is also discussed in this framework.
\end{quote}
\vfill
\begin{center}
December 1993
\end{center}
\vfill

\end{titlepage}

\chapter{Introduction}
Despite many efforts in the last sixty years there does not yet
exist a consistent theory of quantum gravity. Why, then, should
one address an issue like the {\em semi\-classical approximation to
quantum gravity} if there is no theory available
which one could approximate?
The answer is simple. According to the correspondence principle,
the notion of a classical spacetime should emerge in an appropriate
limit from quantum theory. Thus, by applying semiclassical considerations
one hopes to get an insight into some of the structures of the full,
elusive,
theory.

Semiclassical ideas have been very important in the history of
quantum theory. The Bohr-Sommerfeld quantization formula
\be \frac{1}{2\pi\hbar} \oint pdx= n+\frac{1}{2}, \lb{1.1} \ee
for example, was used before the advent of quantum mechanics as
an ad hoc assumption to explain atomic spectra. Formula (1.1) can,
of course, be recovered from quantum mechanics by applying WKB
methods to the Schr\"odinger equation and is even equal to the exact
result in some special cases such as the harmonic oscillator or
the Coulomb potential. Higher order terms in a WKB expansion then
yield corrections to (1.1) of the order of $\lambda/L$, where $\lambda$
is the de Broglie wavelength, and $L$ is the typical dimension of the
system under consideration.

A more impressive application of semiclassical ideas is
Schr\"odinger's approach to his wave equation by using Hamiltonian
methods. These methods had been developed in
the 19th century to provide a joint
formalism of optics and mechanics, which was then, however,
 merely regarded to give a {\em formal} analogy without
 physical significance, since there was no obvious relation between
 the velocity of light and the velocity of a mechanical particle.
This changed only with de Broglie who came up with the idea that
particles {\em are} waves. It then seemed natural for Schr\"odinger
to look for a wave equation which yields the Hamilton-Jacobi
equation in some appropriate limit. What he basically did
in his pioneering work (Schr\"odinger, 1926a) was to take the
Hamilton-Jacobi equation\footnote{In this paper he treats the
stationary case. The time-dependent Schr\"odinger equation was
introduced only in his last paper on {\em quantization as an
eigenvalue problem} (Schr\"odinger, 1926b).}

\be H\left(q_i,\frac{\partial S}{\partial q^i}\right)
\equiv \frac{1}{2m}(\nabla S)^2 +V =E \lb{1.2} \ee

and make the following ansatz for a wave function:

\be \psi\equiv \exp\left(iS/\hbar\right). \lb{1.3} \ee

(The constant $\hbar$ was first left open and later determined
from the spectrum
of the hydrogen atom.) Calculating the second derivatives
of $\psi$ with respect to the $q_i$ and neglecting the second
derivatives of $S$ compared to its first derivatives (this
is motivated by geometrical optics where $S$ varies very little
with $x$) he found after insertion into (1.2) the famous
Schr\"odinger equation

\be \left(-\frac{\hbar^2}{2m}\nabla^2 +V\right)\psi =E\psi.
 \lb{1.4} \ee

In contrast to de Broglie, Schr\"odinger has only waves, but they are
waves in {\em configuration space} and {\em not} in three-dimensional
space like light waves. Semiclassical considerations of this kind
will also be the general theme in this article, but instead of starting
from the Hamilton-Jacobi equation we will accept a "model equation"
for the wave functionals in quantum gravity and then discuss consequences
from the semiclassical expansion of this equation.

To present a last example of the importance of semiclassical considerations
in the history of quantum theory, I would like to write down the
corrections to the classical Maxwell action through quantum
fluctuations of electrons and positrons, which was found
by Heisenberg and Euler (1936) before the development
of QED. In the weak field approximation the corrected Maxwell
Lagrangian reads

\be {\cal L} =\frac{1}{8\pi} ({\bf E}^2- {\bf B}^2)
 + \frac{e^4\hbar}{360\pi^2m^4c^7} \left[
 ({\bf E}^2- {\bf B}^2)^2 +7({\bf EB})^2\right]. \lb{1.5} \ee

This result can of course be found from full QED by "integrating out"
the fermions and expanding in powers of $\hbar$. Since this is
equivalent to an expansion in the number of loops, (1.5) is also
called the one-loop effective Lagrangian. We will encounter an
analogous level of approximation in the case of quantum gravity below.
The knowledge of (1.5) allows the computation of physical
processes such as the scattering of light by external fields
(Delbr\"uck scattering), which has been experimentally observed.

Semiclassical methods have also become increasingly popular
in the field of chaos where people have even coined the term
"postmodern quantum mechanics" for these methods.

What is the situation in quantum gravity? It would be ideal to start
with a "theory of everything" like superstring theory which
encompasses all interactions including gravity in a single quantum
framework. Unfortunately, string theory has not yet reached the
stage where any genuine quantum gravity predictions such as
black hole evaporation can be made. A great deal of work has
therefore been done on the level of {\em quantum general relativity},
 the topic of this volume,
where formal quantization rules are applied to general relativity
without invoking any scheme of unification of interactions.
Whether quantum general relativity exists as a consistent
theory and whether it can be derived from an underlying
"theory of everything" in an appropriate low-energy limit
is an open issue (it depends in particular on the ratio of, say,
the string scale to the Planck scale). As long as the status of
a possible fundamental theory is unknown, it seems perfectly
justified to investigate whether quantum general relativity
can serve as a consistent theory of quantum gravity.
The perturbative non-renormalisability of this approach
may thus not necessarily present a fundamental problem.
(In the following we will always use the term "quantum
gravity" in this sense.)
 While most
of the other contributions in this volume are concerned with the
fundamental level itself, this article will basically
address the following questions. Firstly, how can the framework of
quantum field theory in a given background spacetime be recovered,
 at least formally, from quantum gravity and secondly, can one
calculate corrections to this framework which come from
quantum gravity and may even be testable? Within the framework of
quantum field theory in a given spacetime one can derive
concrete results like the Hawking temperature for black holes.
It would thus be interesting to know the corrections to such
results from quantum gravity.

The organisation of this article is as follows. In the next section
I show how the functional Schr\"odinger equation for matter fields
can formally be recovered from quantum gravity. I also present
the nonrelativistic limit of the Klein-Gordon equation as a
useful formal analogy.

Section 3 is a major part of this contribution and is devoted
to the derivation of correction terms to the Schr\"odinger equation
from quantum gravity. Applications include the corrections to
energy expectation values of scalar fields in de Sitter space
and non-unitary contributions to the evaporation of black holes.
I also make a comparison with an analogous approximation scheme
in QED.

In section 4 I address the issues of decoherence and back reaction,
which are important concepts in the semiclassical expansion.
I briefly discuss examples from quantum mechanics for illustration
and then turn to the analogous situation in quantum gravity.
The modification of semiclassical time through back reaction
is carefully examined.

The last section gives a brief summary
and a brief account of the topics
which have not been addressed in this article.
This includes the relevance of semiclassical
considerations for the problem of time in quantum gravity as well
 as a
comparison with the effective action approach and standard
perturbation theory.

\chapter{Derivation of the Schr\"odinger equation from
         quantum gravity}

The central equation of canonical quantum gravity is the constraint
equation $H\Psi=0$. This was discussed in detail in other
contributions to this volume. In the following we will mainly use
the geometrodynamical language, since semiclassical discussions
are most transparent in this picture. We will comment, however,
on the situation using Ashtekar's variables at the end of this section.

Starting point is the full Wheeler-DeWitt equation
\be {\cal H}\Psi[h_{ab},\phi] \equiv
    \left(-\frac{16\pi G\hbar^2}{c^2}G_{abcd}
    \frac{\delta^2}{\delta h_{ab}\delta h_{cd}}
    -\frac{c^4}{16\pi G}\sqrt{h}(R-2\Lambda)
    +{\cal H}_m\right)\Psi=0, \lb{2.1} \ee

where
$h_{ab}$ is the three-metric, $R$ the three-dimensional Ricci scalar,
$\Lambda$ the cosmological constant,
and ${\cal H}_m$ is the Hamiltonian density for non-gravitational
fields, denoted symbolically below by $\phi$. In the remaining part of this
article we will deal for simplicity exclusively with scalar fields.
There is of course the usual factor ordering ambiguity in (2.1),
but for definiteness we use the naive factor ordering with no first
metric derivatives in the following discussion and comment on the
general case briefly at the end.

One now introduces the parameter
\be M\equiv\frac{c^2}{32\pi G}, \lb{2.2} \ee
with respect to which the semiclassical expansion will be performed.
The Whee\-ler-DeWitt equation then reads

\be    \left(-\frac{\hbar^2}{2M}G_{ab}
    \frac{\delta^2}{\delta h_{a}\delta h_{b}}
    +MV
    +{\cal H}_m\right)\Psi=0, \lb{2.3} \ee

where $V$ stands for $-2c^2\sqrt{h}(R-2\Lambda)$, and we have
introduced a condensed notation, labeling three-metric coefficients by
$h_a$ and components of the DeWitt-metric by $G_{ab}$, since
contractions always involve index pairs.

An expansion with respect to the large parameter $M$ should
lead to sensible results
 if the relvant mass scales of non-gravitational fields
are much smaller than the Planck mass. If there were no
non-gravitational fields, an $M$ expansion would be fully
equivalent to an $\hbar$ expansion and thus to the usual
WKB expansion for the gravitational field (this can be seen from
(2.1) after multiplication with $G$). In the presence of
non-gravitational fields, the $M$ expansion is analogous
to a Born-Oppenheimer expansion, where the large mass of the nuclei
is replaced by $M$, and the small electron mass is replaced by
the mass-scale of the non-gravitational field.

The first subsection is thus devoted to a brief review of the
Born-Oppenheimer approximation in molecular physics. Most of the
steps in that scheme will simply be extrapolated, in a formal way,
to quantum gravity. We then present another helpful analogy --
the nonrelativistic expansion of the Klein-Gordon equation.
Subsection~3 will then present a detailed discussion of the derivation
of the Schr\"odinger equation
from the Wheeler-DeWitt equation, while the last section will give a brief
review of the analogous situation in the Ashtekar framework.

\section{Born-Oppenheimer approximation}

 Consider the following quantum mechanical Hamiltonian
 \be  H=\frac{P^2}{2M} +\frac{p^2}{2m} +V(R,r)
      \equiv \frac{P^2}{2M} +h,  \lb{2.4} \ee
 where $M\gg m$. We seek an approximate solution to the stationary
 Schr\"odinger equation
 \be H\Psi=E\Psi. \lb{2.5} \ee
 We assume that the spectrum of the "light" particle is known for
 each configuration $R$ of the "heavy" particle, i.e.
 \[ h\vert n;R\rangle =\epsilon(R)\vert n;R\rangle. \]
  The full state $\Psi$ is then expanded into these eigenfunctions,
\be \Psi= \sum_k \psi_k(R)\vert k;R\rangle. \lb{2.6} \ee
If one inserts this ansatz into (2.5) one finds, after multiplication
with $\langle n;R\vert$ from the left and using the
fact that these states are orthonormalised,
 an equation for the wave functions $\psi_n$,
\bea & & \left(-\frac{\hbar^2}{2M}\nabla^2_R +\epsilon_n(R)
       -E\right)\psi_n(R) = \nonumber\\
     & & \ \frac{\hbar^2}{M}\sum_k \langle n;R\vert\nabla_R k;R\rangle
      \nabla_R\psi_k(R) + \frac{\hbar^2}{2M}\sum_k
      \langle n;R\vert\nabla_R^2 k;R\rangle\psi_k(R). \lb{2.7} \eea
Note that this is still an exact formula. The Born-Oppenheimer
approximation now consists in the {\em neglection of off-diagonal terms}
in (2.7).The result is an autonomous equation for the $\psi_n$,
which is conveniently written as follows
\bea & & \left(\frac{1}{2M}(-i\hbar\nabla -\hbar A(R))^2
    -\frac{\hbar^2 A^2}{2M}\right. \nonumber\\
   & & \ \ \left. -\frac{\hbar^2}{2M}\langle n;R\vert \nabla^2_R n;R\rangle
     +\epsilon_n(R)\right)\psi_n(R)=E\psi_n(R), \lb{2.8} \eea
 where we have introduced the "Berry connection"
 \be A(R)=-i\langle n;R\vert\nabla_R n;R\rangle. \lb{2.9} \ee
 Note that the momentum of the "slow" particle has been shifted according
 to
 \be P \to P-\hbar A. \lb {2.10} \ee
 In most textbook treatments this connection is not taken into
 account, since it is assumed that the "fast" eigenfunctions can be chosen
 to be real, in which case
 the connection $A$ vanishes. There are, however, situations
 where this cannot be done and $A$ has to be taken into account, see
 for example Jackiw (1988a), and Wudka (1990). As we will see below,
 the complex nature of the "fast" wave functions is essential in quantum
 gravity.

We emphasise that the essential approximation in this scheme consists
in the neglection of the off-diagonal terms in (2.7) and thus the reduction
of the superposition (2.6) to a single product state. As I will
briefly review in section~4 below, this can be justified by
taking the unavoidable interaction with environmental degrees of freedom
into account -- the various components in (2.6) {\em decohere}
from each other. Moreover, this interaction leads to decoherence
{\em within} one component. The position of heavy molecules, for example,
is "measured" by radiation (Joos and Zeh, 1985). In
this sense one can therefore justify the
 substitution of a {\em fixed} position $R_0$ of the "heavy" particle
into the "fast" states $\langle n;R\vert$. It is only in this limit
that molecules with a well-defined shape emerge.

\section{Non-relativistic limit of the Klein-Gordon equation}

The Klein-Gordon equation reads, in the case of vanishing external
field,
\be \left(\frac{\hbar^2}{c^2}\frac{\partial^2}{\partial t^2}
    - \hbar^2\nabla^2 +m^2c^2\right)\varphi({\bf x},t)=0.
     \lb{2.11} \ee
On comparison with (2.3) we recognise that the limit $M\to\infty$
is equivalent to the limit $c\to\infty$ in the Klein-Gordon case.
It is, however, emphasised that this analogy is only a partial
one. The Klein-Gordon equation (2.11) is an equation for a
one-particle wave function, whereas the Wheeler-DeWitt equation (2.3)
is already a "second-quantized" equation, i.e. the corresponding
state is a wave {\em functional} on a configuration space,
whose points are {\em field} configurations. Neverthe\-less, both equations
 have the structure of a wave equation and exhibit many formal
 similarities.

The non-relativistic expansion starts by writing the wave function
$\varphi({\bf x},t)$ as
\be
    \varphi({\bf x},t) = \exp\left(iS({\bf x},t)/\hbar\right) \lb{2.12}
    \ee
and expanding
\be S=c^2S_0 +S_1 +c^{-2}S_2 +\ldots . \lb{2.13} \ee
Inserting this into (2.11) and comparing equal powers of the expansion
parameter $c^2$ yields at order $c^4$
\be (\nabla S_0)^2 =0, \lb{2.14} \ee
so that $S_0$ depends on $t$ only. The next order, $c^2$, yields
\be -\left(\frac{\partial S_0}{\partial t}\right)^2 +m^2
    =0. \lb{2.15} \ee
This is a Hamilton-Jacobi type of equation, which gives real solutions
if $m^2\ge0$, i.e. if there are no tachyons. The solutions can then
immediately be written down:
\be S_0=\pm mt +\ constant. \lb{2.16} \ee
At this order of approximation we thus have for the wave function
\be \varphi_{\mp}({\bf x},t) \approx \exp\left(\pm imc^2t/\hbar
    \right). \lb{2.17} \ee
 Its interpretation is obvious: It describes a particle at rest with
 positive energy (lower sign) or with negative energy (upper sign).
 To neglect superpositions like $\varphi_+ +\varphi_-$ reflects
 the fact that field theoretic effects like pair creation are
 assumed to be
 negligible. Note that both $\varphi_+$ and $\varphi_-$ are complex,
 despite the real character of the Klein-Gordon equation (2.11).
 We assume in the following the use of the positive energy wave
 function.

The next order ($c^0$) yields an equation for $S_1$:
\be 2m\dot{S_1} +(\nabla S_1)^2 -i\hbar\nabla^2 S_1=0.
    \lb{2.18} \ee
This can be simplified by defining the wave function
\be \chi\equiv \exp\left(iS_1/\hbar\right), \lb{2.19} \ee
which leads, for non-vanishing mass, to
\be i\hbar\dot{\chi} =-\frac{\hbar^2}{2m}\nabla^2\chi. \lb{2.20} \ee
At this order one has thus recovered the non-relativistic
Schr\"odinger equation. If one proceeds beyond this order,
one finds, at order $c^{-2}$, the first relativistic corrections.
The equation involving $S_2$ reads
\be i\hbar\ddot{S_1} -i\hbar\nabla^2 S_2 -\dot{S_1}^2
    +2m\dot{S_2} +2\nabla S_1\nabla S_2=0, \lb{2.21} \ee
which can be simplified if one rewrites $S_1$ in terms of $\chi$
and introduces the definition
\be \psi\equiv \chi\exp\left(iS_2/\hbar c^2\right). \lb{2.22} \ee
The wave function $\psi$ then obeys the modified Schr\"odinger equation
\be i\hbar\dot{\psi}= -\frac{\hbar^2}{2m}\nabla^2\psi
    -\frac{\hbar^4}{8m^3c^2}\nabla^2\nabla^2\psi. \lb{2.23} \ee
 The last term on the right-hand side of this equation is, of course,
 just the first relativistic correction to the kinetic energy,
 which can easily be obtained by expanding the square root
 $\sqrt{m^2c^4+p^2c^2}$. The present expansion scheme has, however,
 the merit of systematically producing correction terms also
 in the presence of external fields. In the case of coupling to an
 external gravitational field, some of these correction terms
 are imaginary and thus explicitly violate unitarity in the matter
 sector (L"mmerzahl, 1993). We will encounter similar terms in the
 case of the Wheeler-DeWitt equation, see section~3.

I emphasise that (2.23) does not make sense physically as a differential
equation (the effective Hamiltonian, for example, is unbounded from
below). It only makes sense if the additional term is used to
calculate small corrections to results found on the previous level
of approximation. One can, for example, calculate the
relativistic corrections to the
energy eigenvalues of hydrogen-like pionic atoms:
\bea \Delta E_{rel} &=& -\int d^3x \psi^*_{nlm}
    \left(\frac{\hbar^4}{8m^3c^2}\nabla^2\nabla^2\right)
    \psi_{nlm} \nonumber\\
    & = & -\frac{mc^2}{2}(Z\alpha)^4
      \left(\frac{1}{n^3(l+\frac{1}{2})} -\frac{3}{4n^4}\right).
      \lb{2.24} \eea
We now turn to the interesting case of the Wheeler-DeWitt equation.

\section{Derivation of the functional Schr\"odinger equation
         from the Wheeler-DeWitt equation}

We now make an expansion of the Wheeler-DeWitt equation (2.3) with
respect to $M$, which is formally similar to the non-relativistic
expansion of the Klein-Gordon equation presented above.
To be specific, we take ${\cal H}_m$ to be the Hamiltonian
density of a minimally coupled scalar field, i.e.,
\be {\cal H}_m = \frac{1}{2}\left(-\frac{\hbar^2}{\sqrt{h}}
    \frac{\delta^2}{\delta\phi^2} +\sqrt{h}h^{ab}\phi_{,a}\phi_{,b}
    +\sqrt{h}(m^2\phi^2 +U(\phi))\right), \lb{2.25} \ee
where $U(\phi)$ can be an arbitrary potential. In analogy to
(2.12) we make the ansatz
\be \Psi\equiv\exp\left(iS/\hbar\right), \lb{2.26} \ee
and expand
\be S= MS_0 +S_1+M^{-1}S_2 +\ldots . \lb{2.27} \ee
As remarked above, this expansion scheme will distinguish {\em all}
gravitational degrees of freedom (as being "semiclassical") from the
remaining ones (which are "fully quantum"). One would,
of course, expect that not all degrees of the gravitational field
are semiclassical, such as the "gravitons." It is
straightforward to generalise this scheme to incorporate the more
general situation. Vilenkin (1989), for example, makes the ansatz
\[ h_{ab}= h_{ab}^{(0)} +\sqrt{8\pi G}q_{ab}. \]
Inserting this into the Wheeler-DeWitt equation, one recognises
that the $q_{ab}$ appear on the same footing as $\phi$ and
are therefore "fully quantum."
The expansion is thus, of course, no longer an expansion with respect
to $G$. Generally speaking, the assumption behind this is that the
quantum states of the "fast variables" depend on the "semiclassical
variables" only adiabatically, see e. g. Halliwell and Hawking (1985)
or Kiefer (1987).

In complete analogy to (2.14)ff. one finds after inserting (2.26)
and (2.27) into (2.3) at the highest order, $M^2$,
\be \left(\frac{\delta S_0}{\delta\phi}\right)^2 =0, \lb{2.28} \ee
i.e. that $S_0$ depends only on the three-metric. If many matter
fields are present one can draw this conclusion only if their
kinetic energies are positive (provided, of course, $S_0$
is assumed to be either real or pure imaginary). Since the gravitational
kinetic energy is indefinite, one is thus not able to derive the
opposite limit, where all gravitational degrees of freedom are
quantum and the non-gravitational ones semiclassical.

The next order, $M^1$, yields
\be \frac{1}{2}G_{ab}\frac{\delta S_0}{\delta h_a}
     \frac{\delta S_0}{\delta h_b} +V(h_a) =0. \lb{2.29} \ee
This is the Hamilton-Jacobi equation for the gravitational
field (Peres, 1962). It is equivalent to all ten of Einstein's
field equations
 (DeWitt, 1967; Gerlach, 1969; DeWitt, 1970). I want to briefly
  review how this
equivalence can be shown, i. e. how the "dynamical laws follow from
the laws of the instant" (Kucha\v{r}, 1992).

Suppose one is given a solution $S_0$ to (2.29). From this one can
immediately read off the geometrodynamical field momentum
\be \pi^{ab}=M\frac{\delta S_0}{\delta h_{ab}}. \lb{2.30} \ee
One half of Hamilton's equations of motion then give the connection
between the field momenta and the "velocities" $\dot{h_{ab}}$,
\be \dot{h_{ab}} =-2N K_{ab} +N_{a\vert b} +N_{b\vert a}, \lb{2.31} \ee
where $K_{ab}$ are the components
of the extrinsic curvature of three-space, whose
relation to the momenta is
\be K_{ab}=-\frac{16\pi G}{c^2}G_{abcd}\pi^{cd}, \lb{2.32} \ee
 and
$N$ and $N_a$ are respectively the lapse function and the shift vector,
for both of which a choice must be made in order for $\dot{h_{ab}}$
to be specified. After this has been done,
and an "initial" three-geometry has been chosen,
 (2.31) can be integrated
to yield a whole {\em spacetime} with a definite foliation and a definite
choice of coordinates on each member of this foliation.
 One can thus combine all
the trajectories in superspace which describe the same spacetime
to a "sheaf" (DeWitt, 1970). Lapse and shift can be chosen in such
a way that these curves comprise a "sheaf of geodesics." Superposing
many WKB solutions of the gravitational field, Gerlach (1969)
has demonstrated how {\em one} specific trajectory in configuration
space, i. e. one specific spacetime, can be traced out by
a wave packet.

Since $S_0$ satisfies, in addition to the Hamilton-Jacobi equation
(2.29), the three equations
\be \left(\frac{\delta S_0}{\delta h_{ab}}\right)_{\vert b}
    =0, \lb{2.33} \ee
 the "complete solution" depends on two arbitrary functions on three-space.
 The relations (2.33) follow from an expansion of the momentum
 constraints in powers of $M$. As was shown by Moncrief and
 Teitelboim (1972), these equations follow from the validity
 of the Hamilton-Jacobi equation (2.29) itself, if one takes
 into account the Poisson bracket relations between the
 constraints ${\cal H}$ and ${\cal H}_a$.
 While (2.29) and (2.33) are equivalent to the $G_{00}$ and
 $G_{0i}$ part of the Einstein equations, the remaining six
 field equations can be found by differentiating (2.31) with respect
 to $t$ and eliminating $S_0$ by making use of (2.31) and
 (2.29). Basically, this is the same procedure as can already be
 done in Newtonian mechanics.

The Hamilton-Jacobi equation (2.29), as it stands, is actually an infinity
of equations, one at each space point. An alternative viewpoint
is also useful (Giulini, 1993). It interprets (2.29) as being
multiplied with a "test function" and thus
represents {\em one} equation for {\em each} choice of lapse, i.e.
for each choice of foliation. The advantage of this viewpoint
is that, although it may be impossible to solve the Hamilton-Jacobi
equation for each choice of lapse, it may be possible to do so
for specific choices. This may be enough in certain situations
to draw
interesting
 physical conclusions. One way or the other, it will be assumed
in the following that a specific solution of (2.29) has been
chosen and a specific spacetime with a specific foliation has
been constructed from it.

It has been remarked (Kucha\v{r}, 1992) that the sandwich problem
may present some obstacle in solving the Hamilton-Jacobi equation.
This is because a complete solution to (2.29) can be interpreted
as generating a canonical transformation between two three-geometries
which are assumed to be connected by a single spacetime
(which is in a sense "sandwiched" between these three-geometries).
The uniqueness of the interpolating four-geometry was only proven
in special cases, for example if $2\rho-R>0$ (Belasco and
Ohanian, 1969). It may thus happen that the complete solution
to (2.29) cannot generate a canonical transformation between
all three-geometries. In the semiclassical approximation, however,
we use a {\em special} solution to (2.29) and can always construct
a spacetime from a given three-geometry using the construction
presented above.

Proceeding with the semiclassical expansion, one finds at order
$M^0$ an equation involving also $S_1$:
\be G_{ab}\frac{\delta S_0}{\delta h_a}\frac{\delta S_1}
    {\delta h_b} -\frac{i\hbar}{2}G_{ab}
    \frac{\delta^2 S_0}{\delta h_a\delta h_b}
     +{\cal H}_m =0. \lb{2.34} \ee
As in the Klein-Gordon case, one can simplify this equation
by introducing the wave functional
\be \chi=D[h_a]\exp\left(iS_1/\hbar\right). \lb{2.35} \ee
We will {\em choose} $D$ in such a way that the equations become
simple. This is achieved by demanding
\be G_{ab}\frac{\delta S_0}{\delta h_a}\frac{\delta D}{\delta h_b}
    -\frac{1}{2}G_{ab}\frac{\delta^2 S_0}{\delta h_a\delta h_b}
    D= 0, \lb{2.36} \ee
which can also be written in the form of a "conservation law"
\be G_{ab}\frac{\delta}{\delta h_a}\left(\frac{1}{D^2}
    \frac{\delta S_0}{\delta h_b}\right)=0. \lb{2.37} \ee
The interpretation of this equation becomes immediately clear if one
writes down its pendant in one-dimensional quantum mechanics:
\[ \frac{\partial}{\partial x}\left(\frac{1}{D^2}
    \frac{\partial S_0}{\partial x}\right)=0. \]
 This is nothing but the continuity equation for time-independent
 states.
 The solution of this simple equation gives the well known
 expression for the WKB prefactor $D\propto\sqrt{p}$.

Using (2.37) one finds from (2.34)
\be i\hbar G_{ab}\frac{\delta S_0}{\delta h_a}\frac{\delta\chi}
    {\delta h_b}\equiv i\hbar\frac{\delta\chi}{\delta\tau}
    ={\cal H}_m\chi. \lb{2.38} \ee
 This is the functional Schr\"odinger equation in its local form
 (the Tomonaga-Schwinger equation) for quantum fields propagating
 on the classical spacetimes described by $S_0$. Note that
 the Schr\"odinger equation cannot be recovered if $S_0$ turns
 out to be a constant.

The above approach to semiclassical gravity has been discussed
from various points of view by several authors including
DeWitt (1967), Lapchinsky and Rubakov (1979), Banks (1985),
Halliwell and Hawking (1985), Hartle (1986), and Kiefer (1987).

 In (2.38) one has introduced a time functional $\tau(x;h_a]$
 according to
 \be G_{ab}(x)\frac{\delta S_0}{\delta h_a(x)}
     \frac{\delta\tau(x;h_a]}{\delta h_b(y)}= \delta(x-y).
     \lb{2.39} \ee
 This "many-fingered" time (which is also called
 {\em WKB time}) is defined on configuration space, but also yields
 a time parameter in each of the spacetimes described by $S_0$.
 This is clear from the above construction, since each spacetime
 is constructed from a specific foliation, which is labelled
 by $\tau$. It is for this reason that the "spacetime problem"
 (Isham, 1992; Kucha\v{r}, 1992) does not come into play in the
 semiclassical approximation. This problem refers to the fact that
 $\tau$ is not a spacetime scalar, i. e. that it depends on the
 embedding which is chosen. While this may in fact present a problem
 for approaches which attempt to use functions like $\tau$ as
 fundamental time functions in quantum gravity, there is no such
 ambiguity in the semiclassical approximation, where the spacetimes
 emerge with a specific foliation.

We also note that the hypersurfaces $\tau=constant$ in configuration
space do not coincide with the hypersurfaces $S_0=constant$. This
can be immediately seen by comparing the definition of $\tau$ (2.39)
with the Hamilton-Jacobi equation (2.29).

Another problem discussed in this context is the "global time
problem." As simple minisuperspace models indicate, $\tau$
may not exist globally in the space of three-geometries.
This is, however, not a real obstacle for the semiclassical
approximation in that one can simply restrict the attention
to regions in superspace where it exists. Since we will be interested
in the following to quantum corrections only in the neighbourhood
of a given spacetime this does not really present a restriction.

Barbour (1992) has stressed that the above introduced WKB time
is very similar to the notion of {\em ephemeris time} as it is
used by astronomers. This time is determined in retrospect from
actual observations of the celestial bodies, exploiting the
dynamical equations of the gravitational field. The above WKB
time thus plays a double role: From one point of view one may
simply interpret it as a coordinate on spacetime, but on the other
side it is constructed from the geometry of a given three-dimensional
space and is thus a quantity which in principle can be determined
from astronomical observations. That ultimately the whole
three-geometry is implemeted into the construction of this notion
of time can very clearly be seen in the motion of the binary pulsar
$PSR\ 1913+16$: If one attempts to determine ephemeris time from
this system (exploiting the period decrease due to the radiation
of gravitational waves) one reaches consistency only if the whole
motion of the Galaxy and its "pull" on the binary pulsar is
quantitatively taken into account (Damour and Taylor, 1991).

Up to this order, the total wave functional thus reads
\be \Psi\approx\frac{1}{D}\exp\left(iMS_0[h_{ab}]/\hbar
   \right)\chi[h_{ab},\phi],
     \lb{2.40} \ee
where $\chi$ obeys the Schr\"odinger equation (2.38).

\section{Semiclassical gravity in the connection representation}

Instead of using the three-metric and is conjugate momentum as
the basic variables one can perform a canonical transformation
at the classical level to a set of new canonical variables consisting
of the densitised triad and a complex $SO(3)$ connection.
Details are presented in various other contributions to this
volume. The fundamental equation is, of course, again of the form
$H\Psi=0$, but the Hamiltonian reads differently, see Ashtekar (1991):
  \bea H &=& \frac{G\hbar^2}{4\pi}\epsilon^{ijk}F_{ab}^{k}
  \frac{\delta^{2}}{\delta A^{i}_{a}\delta A^{j}_{b}}
    +\hbar^2\frac{\delta^{2}}{\delta\phi^{2}}
     + 2G^{2}\hbar^2\partial_{a}\phi\partial_{b}\phi\frac{\delta^{2}}
      {\delta A_{a}^{i}\delta A^{i}_{b}} \nonumber\\
    & & \ +\frac{\sqrt{2}\hbar^3}{3}(G^{3}m^{2}\phi^{2}+G^{2}\Lambda)
    \eta_{abc}\epsilon^{ijk}
         \frac{\delta^{3}}{\delta A_{a}^{i}\delta A_{b}^{j}\delta A_{c}^{k}},
          \lb{2.41} \eea
     where$F_{ab}^{k}=\partial_{a}A_{b}^{k}-\partial_{b}A^{k}_{a}-
     \epsilon^{klm}A_{a}^{l}A_{b}^{m}$ is the field strength tensor
      associated with  the complex connection
${\bf A}$, and $\eta_{abc}$ is the metric-independent totally skew-symmetric
 density of weight $-1$. I have used here a {\em rescaled}
 connection $A^i_a\equiv G\bar{A^i_a}$, where $\bar{A^i_a}$ denotes
 the connection as it is used in Ashtekar (1991). Such a rescaling is
 convenient for performing the semiclassical expansion and has also been
 proven useful in the investigation of the strong gravity limit,
 see Ashtekar (1988). The distinguished property of (2.41) is the fact
 that the potential term proportional to $R$ has been "swallowed" by
 the new variables. In addition, it contains only terms with
 functional derivatives (even third order derivatives) and that it is
 intrinsically complex since ${\bf A}$ is complex. Note also that
 a factor of $G$ is associated with each functional derivative
 with respect to $A_a^i$ which comes with a term containing $\phi$.

The semiclassical expansion now proceeds analogously to the
geometrodynamical case of the previous section (the parameter
$M$ is now just the inverse of $G$). The functional $S_0$ does not
depend on non-gravitational fields, and it obeys the
Hamilton-Jacobi equation
\be \frac{\epsilon^{ijk}}{4\pi}F_{ab}^{k}\frac{\delta S_{0}}
{\delta A_{a}^{i}}\frac{\delta S_{0}}{\delta A_{b}^{j}}
+\frac{i\sqrt{2}}{3}\Lambda\eta_{abc}\epsilon_{ijk}
\frac{\delta S_{0}}{\delta A_{a}^{i}}\frac{\delta S_{0}}
{\delta A_{b}^{j}}\frac{\delta S_0}{\delta A_c^k}=0. \lb{2.42} \ee
 Note that since the momentum conjugate to $A_{a}^{i}$, $\tilde{E_{i}^{a}}$,
  is replaced by $\delta/\delta A_{a}^{i}$ (without an $i$)
   in the Schr\"{o}dinger representation, the momentum is given
    by $\tilde{E_{i}^{a}}=i\delta S_{0}/\delta A_{a}^{i}$. The triad
   $\tilde{E_i^a}$ is not necessarily real, but it can be made
   real since it satisfies Gauss's law.
  Note that $S_0=constant$ is always a solution of this equation.

The next order ($G^{0}$) yields the functional Schr\"{o}dinger
 equation for the wave functional $\chi\equiv De^{iS_{1}}$,
\be i\hbar\frac{\epsilon^{ijk}}{4\pi}F_{ab}^{k}\frac{\delta S_{0}}
{\delta A_{a}^{i}}\frac{\delta \chi}{\delta A_{b}^{j}}\equiv
 i\hbar\frac{\delta\chi}{\delta\tau}=\tilde{H_{m}}\chi, \lb{2.43} \ee
where the Hamiltonian density $\tilde{H_{m}}$ is now given by the expression
\bea \tilde{H_{m}} &=& -\frac{\hbar^2}{2}\frac{\delta^{2}}
     {\delta\phi^{2}}+\frac{\delta S_{0}}{\delta A_{a}^{i}}
     \frac{\delta S_{0}}{\delta A_{b}^{i}}\partial_{a}\phi\partial_{b}\phi
 + \frac{i}{3\sqrt{2}}m^2\phi^2 \eta_{abc}\epsilon_{ijk}
 \frac{\delta S_{0}}{\delta A_{a}^{i}}\frac{\delta S_{0}}
 {\delta A_{b}^{j}}\frac{\delta S_{0}}{\delta A_{c}^{k}}\nonumber\\
 & & +\sqrt{2}\Lambda\hbar\eta_{abc}\epsilon_{ijk}\frac{\delta S_{0}}
 {\delta A_{a}^{i}}\left(\frac{\delta S_{0}}{\delta A_{b}^{j}}
 \frac{\delta}{\delta A_c^k}-\frac{1}{D}\frac{\delta S_{0}}
 {\delta A_{b}^{j}}\frac{\delta D}{\delta A_c^k}+\frac{\delta^2S_0}
 {\delta A_b^j\delta A_c^k}\right). \lb{2.44} \eea

A comparison with (2.41) exhibits that
 $\tilde{H_{m}}$ is equal to $H$ as evaluated on the classical
  gravitational background determined by the Hamilton-Jacobi equation
   {\em except} for the last three $\Lambda$- dependent terms in (2.44)
   which arise due to the presence of the third functional derivatives
   in (2.41). These terms can be absorbed by a redefinition
   of the wave functional, but their interpretation is not yet clear.

 It seems that the traditional semiclassical approximation is tied
 to the presence of only second order derivatives (i.e. quadratic
 momenta) in the Hamiltonian. Consider, for example, the anharmonic
 oscillator in quantum mechanics, which is defined by the Hamiltonian
 \[ H=\frac{p^2}{2} +\lambda x^4. \]
 If one wished to use the momentum representation instead of the
 position representation for the wave function in the Schr\"odinger
 equation, one would have to substitute fourth order derivatives
 in the momentum for the second term in $H$. Performing a WKB expansion
 one finds terms analogous to the ones in (2.41), which one would
 have to absorb through a redefinition of the wave function to get
 a transparent interpretation. This problem reflects of course
 the unsolved issue of the Hilbert space structure
 of the full theory (which would allow much less states than
 the number of solutions to $H\Psi=0$).

A second major difference to the geometrodynamical case
is the complexity of the connection, as mentioned above.
One has been able to write (2.43)
  as a functional Schr\"odinger equation after,
  in an appropriate region of configuration space, a time functional
   $\tau(\bf{x};\bf{A}]$ was introduced, which satisfies
\be \delta({\bf x}-{\bf y})=\frac{\epsilon^{ijk}}{4\pi}F^k_{ab}
({\bf y})\frac{\delta S_{0}}{\delta A^i_a({\bf y})}
\frac{\delta\tau({\bf x};{\bf A}]}{\delta A^j_b({\bf y})}. \lb{2.45} \ee
This time functional, however, is complex
and thus not viable as a semiclassical time parameter. Since one
generally assumes that the functionals in the connection
representation are holomorphic, one can take the {\em real part} of $\tau$
as a candidate for WKB time.

Since one wants to recover {\em real} quantum gravity at a certain
stage, one must implement appropriate reality conditions (Ashtekar, 1991).
The assumption, which is motivated by a comparison with the Bargmann
representation for the harmonic oscillator, is that they are used
to determine the inner product which has to be imposed on the
physical states. The present semiclassical approximation, however,
only works on the level of the differential equation and therefore
does not "know" about the Hilbert structure
(see the remarks above). It may even be possible
that both concepts are in conflict with each other, i. e. that
the semiclassical approximation "runs out" of the Hilbert space
(Louko, 1993). It is definitely not possible to implement the reality
conditions on the wave functional
 in the same way as the constraint equation. One may
therefore be able to decide the compatibility only after the
appropriate Hilbert space structure of the full theory has been
determined.

The above derivation has to be contrasted with the derivation of the
Schr"\-ding\-er equation by Ashtekar (1991). He expands the
{\em classical} Hamiltonian constraint up to second order around
some {\em given} classical configuration, which is chosen to be
a flat triad on a non compact manifold.
The important feature of the truncated constraint is its
{\em linearity} in the triad, i. e. in the canonical
momentum.
As in the full theory, quantisation proceeds by implementing
the truncated constraint as a condition on wave functionals. The result is
\bea & & -\hbar\frac{1}{G}(\Delta\frac{\delta}{\delta A^T(x)})
\Psi(A^{TT}, A^L, A^T, A^A)\nonumber\\ & & \;
    =GA^{TTab}(x)\left(A^{TT}_{ab}(x)\right)^*
    \Psi(A^{TT}, A^L,A^T, A^A). \lb{2.46} \eea
The internal indices have here all been converted to space indices
by the use of the given background triad, and we have decomposed the
resulting "connection tensor" into its transverse-traceless,
longitudinal, trace, and antisymmetric parts.

The important step is now to identify the term $G(\Delta)^{-1}A^T(x)$
as a "many-fingered" parameter $\tau(x)$, which has the dimension of time.
 (This is a {\em nonlocal} expression in the connection.)
As in the above semiclassical expansion, it is, however,
complex and thus not suitable as a physical time. But, again, assuming
the holomorphicity of the wave functional one can identify its
{\em imaginary part} as physical time, since then the left-hand side
of (2.46) reads
\be -\hbar\frac{\delta}{\delta\tau(x)}\Psi
   =i\hbar\frac{\delta}{\delta(\mbox{Im}\tau)(x)}
   \Psi, \lb{2.47} \ee
while the right-hand side is the "physical" Hamiltonian in this
order, i. e. it generates the dynamics of the transverse traceless
degrees of freedom (the "gravitons").
One has thus recovered, in the weak field approximation, the
Schr\"odinger equation from quantum gravity.

It was important in the above derivation that the classical
constraint has been expanded up to quadratic order in the deviation
of the fields from their flat space values, since the left-hand side
of (2.46) would have been zero in the first order approximation.
This has already been recognised by Kucha\v{r} (1970) in his analysis
of the weak field limit in the geometrodynamical language: The
linear gravitons do not contain "information about time." This information
shows first up in
the gravitational field which they produce in the next order.
Ashtekar (1991) has related an {\em intrinsic} degree of freedom
in configuration space, the imaginary part
of the trace part of the connection, to
physical time. In the geometrodynamical language, this
imaginary part is basically given by the trace of the extrinsic
curvature and is thus an "extrinsic time."  Not surprisingly,
this was the time parameter chosen by Kucha\v{r} (1970),
who gave a lucid geometric interpretation: Consider a hypersurface
in flat spacetime by specifying $t$ as a function of $x^i$. If this
hypersurface is attained from a flat hypersurface by a small
deformation, Kucha\v{r} showed that the intrinsic geometry
is only sensitive to the second order in this deformation
($h_{ij}\approx\delta_{ij}+t,_it,_j$), while the extrinsic geometry
is already sensitive to the first order ($K_{ik}\approx t,_{ik}$)
and thus seems to be more appropriate in defining a time variable.

What are the main differences of this approach to the semiclassical
expansion, which was presented above?
One difference is that, whereas above the whole gravitational field
was semiclassical in the first order, in the Ashtekar approach
some of the degrees of freedom are treated as fully quantum, i. e.
they appear in the Hamiltonian in the Schr\"odinger equation.
This is, however, not a major difference since, as has been remarked
in the previous section, one could easily generalise the scheme
to incorporate such a situation by expanding with respect to
a parameter different from $G$. It is thus not surprising that
Ashtekar's definition of time contains the gravitational constant.

What is more important is that in the second approach a classical
background has been chosen {\em ab initio} and only a truncated
constraint, being linear in the canonical momentum, has been
quantised. It is far from clear whether the same results
can be recovered from the full theory in an appropriate
limit.

Whereas time in the semiclassical approximation is constructed
from the phase of the WKB state and thus essentially
state-dependent, the weak field approach constructs time always
from the same part of the configuration variables and is thus
state-independent. Moreover, the semiclassical limit seems to be
inappropriate for studying the weak field limit, since a flat
background spacetime in its standard foliation corresponds to
$S_0=constant$ ("no change of configuration" $\leftrightarrow$
"no WKB time") and thus does not allow the recovery of a
Schr\"odinger equation in this case. This difficulty of incorporating
"flat spacetime" into the picture of quantum geometrodynamics
will again show up in the next order of this approximation
scheme.

\chapter{Corrections to the Schr\"odinger equation
         from quantum gravity}
 \section{General derivation and discussion}

We continue with the semiclassical expansion scheme to the
order $M^{-1}$, which leads to the calculation of
{\em correction terms} to the functional Schr\"odinger equation
(2.38). This order yields an equation involving $S_2$
(compare (2.27)) and reads explicitly
\bea & & G_{ab}\frac{\delta S_0}{\delta h_a}\frac{\delta S_2}
     {\delta h_b}  + \frac{1}{2}G_{ab}\frac{\delta S_1}
  {\delta h_a} \frac{\delta S_1}{\delta h_b} -\frac{i\hbar}{2}
     G_{ab}\frac{\delta^2S_1}{\delta h_a\delta h_b}\nonumber\\
    & & \; +\frac{1}{\sqrt{h}}\left(\frac{\delta S_1}{\delta\phi}
    \frac{\delta S_2}{\delta\phi} -\frac{i\hbar}{2}
    \frac{\delta^2 S_2}{\delta\phi^2}\right)=0. \lb{3.1} \eea
One first rewrites this equation by substituting $S_1$
in favour of $\chi$, see (2.35), and then makes the following ansatz
for $S_2$:
\be S_2[h_a,\phi]= \sigma_2[h_a] +\eta[h_a,\phi]. \lb{3.2} \ee
Eq. (3.1) will be simplified,
if one demands that $\sigma_2$ obeys:
\be G_{ab}\frac{\delta S_0}{\delta h_a}\frac{\delta\sigma_2}
     {\delta h_b} -\frac{\hbar^2}{D^2}G_{ab}\frac{\delta D}
     {\delta h_a}\frac{\delta D}{\delta h_b}
     +\frac{\hbar^2}{2D}G_{ab}\frac{\delta^2 D}{\delta h_a
     \delta h_b}=0. \lb{3.3} \ee
The interpretation of this equation becomes immediately clear,
if one writes out its analogue in one-dimensional quantum mechanics:
\be \sigma_2'= -\frac{\hbar^2}{4}\frac{p''}{p^2}
   +\frac{3\hbar^2}{8}\frac{p'^2}{p^3}, \lb{3.4} \ee
where, again, $p=\partial S/\partial x\equiv S'$.
Eq. (3.4) is nothing but the equation for the second-order
WKB correction, see e. g. Landau and Lifshitz (1975).
(Recall that for pure gravity an expansion in $G$ is fully
equivalent to an expansion in $\hbar$.) The separation of $\sigma_2$
from $S_2$ thus serves to separate the pure gravitational
WKB(2) factor (which in this context is uninteresting) from
the relevant part, which satisfies the corrected Schr\"odinger
equation. We then find from (3.1) an equation involving $\eta$ only:
\bea G_{ab}\frac{\delta S_0}{\delta h_a}\frac{\delta\eta}
     {\delta h_b} &=& \frac{\hbar^2}{2\chi}
     \left(-\frac{2}{D}G_{ab}\frac{\delta\chi}{\delta h_a}
     \frac{\delta D}{\delta h_b} +G_{ab}\frac{\delta^2\chi}
     {\delta h_a\delta h_b}\right)\nonumber\\
     & & + \frac{i\hbar}{\sqrt{h}\chi}\frac{\delta\eta}
     {\delta\phi}\frac{\delta\chi}{\delta\phi}
     +\frac{i\hbar}{2\sqrt{h}}\frac{\delta^2\eta}
     {\delta\phi^2}. \lb{3.5} \eea
 Up to this order, the total wave functional is thus of the form
 \be \Psi=\frac{1}{D}\exp\left(\frac{i}{\hbar}(MS_0+
     \sigma_2M^{-1})\right)
     \chi\exp\left(\frac{i\eta}{M\hbar}\right). \lb{3.6} \ee
 The wave functional
 \be \psi\equiv
     \chi\exp\left(\frac{i\eta}{M\hbar}\right) \lb{3.7} \ee
 then obeys the "corrected Schr\"odinger equation" (Kiefer and Singh, 1991)
 \be i\hbar\frac{\delta\psi}{\delta\tau}
     = {\cal H}_m\psi +\frac{h^2}{M\chi}G_{ab}
     \left(\frac{1}{D}\frac{\delta D}{\delta h_a}
     \frac{\delta\chi}{\delta h_b}-
     \frac{1}{2}\frac{\delta^2\chi}{\delta h_a\delta h_b}
     \right)\psi. \lb{3.8} \ee
 Thus, if one knows the solutions to the previous order equations,
 one can evaluate the correction term on the right-hand side of (3.8).
 Note that it is {\em not} sufficient to know the solution, $\chi$,
 of the Schr\"odinger equation (2.38), but that the knowledge of the
 gravitational prefactor, $D$, is also required.

 As in the case of the uncorrected Schr\"odinger equation,
 Eq. (3.8) is equivalent to its integrated form, since a specific
 slicing of a specific spacetime has been chosen upon giving
 a solution to the Hamilton-Jacobi equation (2.29).

 To write (3.8) in a more transparent way, we decompose the
 vector fields on the right-hand side of (3.8) into their
 respective components tangential and normal to the integral curves
 of the vector field $G_{ab}\delta S_0/\delta h_a$. This will
 enable us to use the Schr\"odinger equation (2.38) for the tangential
 part. We first decompose the first derivative according to
 \be G_{ab}\frac{\delta\chi}{\delta h_a}=
     \alpha\ G_{ab}\frac{\delta S_0}{\delta h_a} +a_b, \lb{3.9} \ee
 where
 \[ \frac{\delta S_0}{\delta h_a} a_a=0. \]
 The coefficient $\alpha$ is determined by multiplying each side
 of (3.9) with $\delta S_0/\delta h_b$, summing over $b$, and
 making use of the Hamilton-Jacobi equation (2.29) as well as
 the Schr\"odinger equation (2.38). One finds
 \be \alpha=\frac{i}{2\hbar V}{\cal H}_m\chi. \lb{3.10} \ee
 This, of course, holds only if the potential $V$ is non-vanishing,
 i. e. in regions where the vector field $G_{ab}\delta S_0
 /\delta h_a$ is not "lightlike." (Note that in the general case
 $V$ is nonzero, since it also includes contributions from
 "macroscopic" matter sources.) In the "lightlike" case $\alpha$
 cannot be unambiguously determined, since the addition of a
 "lightlike" vector on the right-hand side of (3.9) does not change
 the scalar product of the left-hand side with $\delta S_0/\delta h_b$.
 But it is only this scalar product which
  is determined by the previous order equation
 (2.38).

I emphasise that the component $a_b$ is {\em not} determined
by the previous order equations, but only by the boundary
conditions which are imposed on the {\em full} wave functional
$\Psi$. As can be explicitly seen from (3.8), the second derivatives
with respect to the three-metric now come into play.
We will assume in the following that these component can be neglected,
i. e. we will choose the
boundary conditions such that $\chi$ is "peaked" around the
considered worldline in configuration space, i. e. along
the particular spacetime we have chosen (Cosandey, 1993).

The second derivative term on the right-hand side of (3.8) is
now also decomposed into its tangential and normal components,
and use of the previous order equations as well as of (3.10)
is made. After some calculation one can then put (3.8) in
the form (Kiefer and Singh, 1991)
\be i\hbar\frac{\delta\psi}{\delta\tau}= {\cal H}_m\psi
    +\frac{4\pi G}{c^4\sqrt{h}(R-2\Lambda)}
    {\cal H}^2_m\psi +i\hbar\frac{4\pi G}{c^4}
    \frac{\delta}{\delta\tau}\left(\frac{{\cal H}_m}
    {\sqrt{h}(R-2\Lambda)}\right)\psi. \lb{3.11} \ee
This is the central equation of this section. Several comments
are in order.

(1) The WKB(1) prefactor $D$ has disappeared from the corrected
Schr\"odinger equation after use of (2.36) was made. This is fortunate,
since we do not have to deal with the pure gravitational part
of the full wave functional.

(2) As remarked above, this form of the corrected Schr\"odinger
equation only holds for non-vanishing gravitational potential.
In the general case one must resort to the form (3.8).

(3) The correction terms in (3.11) are, of course, only formal
as long as no regularisation scheme is brought into play to
cure the divergences which come together with the functional
derivatives. We will show in the section on applications how
concrete physical prediction can nevertheless be extracted
from these terms in special cases.

(4) The terms in (3.11) are independent of the factor ordering
which is chosen for the gravitational kinetic term in (2.1).There
may, however, be an ambiguity in the factor ordering of the
gravity-matter coupling.

(5) If "macroscopic" matter sources are present, one has to
substitute $\sqrt{h}(R-2\Lambda)$ in the denominator of (3.11) by
$\sqrt{h}(R-2\Lambda)-\epsilon_m/2Mc^2$, where $\epsilon_m$ is
an effective matter energy density. That $M\propto G^{-1}$
comes explicitly into play in this expression, does not
present any problem as long as the corresponding correction terms
are small. This will become important in the example of
black hole evaporation discussed below.

(6) In addition to the Wheeler-DeWitt equation (2.1) one must
also expand the momentum constraints. This is straightforward and
leads to the result that the wave functional remains unchanged
under a spatial diffeomorphism {\em at each order} of this approximation
scheme.

(7) The second correction term in (3.11) is pure imaginary and thus
leads to a {\em violation of unitarity}. The occurrence of such a term
is not surprising, since we have attempted to write down an
effective equation for the non-gravitational fields alone.
This can also be recognised from the point of view of the
conservation laws for the wave functional. The full Wheeler-DeWitt
equation (2.1) is of the Klein-Gordon type and thus obeys the
following conservation law:
\be \frac{1}{M}G_{ab}\frac{\delta}{\delta h_a}
    \left(\Psi^* \stackrel{\leftrightarrow}
    {\frac{\delta}{\delta h_b}}\Psi\right)
    +\frac{1}{\sqrt{h}}\frac{\delta}{\delta\phi}
    \left(\Psi^* \stackrel{\leftrightarrow}
    {\frac{\delta}{\delta\phi}}\Psi\right)=0. \lb{3.12} \ee
Applying the $M$-expansion for the wave functional to this
conservation law, one finds at the level of the corrected Schr\"odinger
equation
\bea & & \frac{\delta}{\delta\tau}(\psi^*\psi)
     +\frac{\hbar}{2i\sqrt{h}}\frac{\delta}{\delta\phi}
     \left(\psi^*\stackrel{\leftrightarrow}
     {\frac{\delta}{\delta\phi}}\psi\right) \nonumber\\
    & & \ -\frac{i\hbar}{2M}G_{ab}
    \left[\frac{\delta}{\delta h_a}\left(\psi^*
    \stackrel{\leftrightarrow}{\frac{\delta}{\delta h_b}}
    \psi\right) -\frac{1}{D}\frac{\delta D}{\delta h_a}
    \psi^*\frac{\delta}{\delta h_b}\psi
    \right] =0. \lb{3.13} \eea
The first two terms together are the "Schr\"odinger current"
connected with (2.38), while
the remaining terms yield correction terms proportional to
$M^{-1}$. Upon functionally integrating this equation over the field $\phi$
and making the standard assumption that $\psi$ falls off for large
field configurations, one finds from (3.13)
\be \frac{d}{dt}\int{\cal D}\phi\ \psi^*\psi
   =\frac{8\pi G}{c^4}\int{\cal D}{\phi}
   \int d^3x\ \psi^*\frac{\delta}{\delta\tau}
   \left(\frac{{\cal H}_m}{\sqrt{h}(R-2\Lambda)}
   \right)\psi. \lb{3.14} \ee
A comparison with (3.11) immediately exhibits that the violation
of the Schr"\-ding\-er conservation law produces the imaginary correction
term in (3.11). I note again that a similar term emerges in the
nonrelativistic expansion of the Klein-Gordon equation in
the presence of an external Newtonian potential (L\"ammerzahl, 1993).

Corrections to the Schr\"odinger equation have been discussed,
in the context of one-dimensional minisuperspace models,
by a number of authors including Brout and Venturi (1989), Singh (1990),
Kowalski-Glikman and Vink (1990), and Bertolami (1991) (see also
Singh (1993) for a review). The above, systematic, expansion in the
context of the full Wheeler-DeWitt equation has been discussed by
Kiefer and Singh (1991).

The above semiclassical expansion scheme can also be applied
to standard quantum field theories like QED, see Kiefer,
Padmanabhan, and Singh (1991), and Kiefer (1992a). The purpose of
such an application lies primarily in the test of its viability within
a well-understood framework. It provides, however, also a
useful insight into the semiclassical expansion of QED itself.
It demonstrates, for example, that one can start from the
{\em stationary} equation for full QED and derive, as in the
gravitational case, a {\em time-dependent} Schr\"odinger equation
for the matter fields in an external electromagnetic background.
Again, the "fast" matter degrees of freedom follow the "slow"
electromagnetical degrees of freedom, which can thus serve
as an "intrinsic clock."

Formally, the expansion is performed with respect to the electric
charge (after a suitable rescaling of the vector potential has been
made). The expression for the full wave functional up to order
$e^0$ then reads, in analogy to the gravitational expression
(2.40),
\be \Psi\approx \frac{1}{D}\exp\left(\frac{iS_0[{\bf A}]}{\hbar e^2}
    \right)\psi[{\bf A}, \phi,\phi^{\dagger}], \lb{3.15} \ee
where ${\bf A}$ is the vector potential, and $\phi$ is a charged
scalar field. The functional $\psi$ obeys a functional Schr\"odinger
equation, in which time is defined by $S_0$.
Corrections to this Schr\"odinger equation can be found in order
$e^2$ and look very similar to the gravitational terms in (3.11).
In QED they can be easily interpreted in terms of Feynman diagrams
with an internal photon line, see Kiefer (1992a). This in turn
provides a bridge to the standard expansion of the effective action.

I want to conclude this section with a brief remark on the
correction terms in the Ashtekar variables. Basically, the expansion
proceeds as in the geometrodynamical case. The main difference is that
there is {\em no} $R$ term in the analogue of (3.11), since the potential
$V$ in the Ashtekar case is given by the cosmological constant
 and the macroscopic
matter contribution only.

\section{Applications}
\subsection{A minisuperspace example}

The simplest example is to neglect all inhomogeneous degrees
of freedom and keep only homogeneous ones like the scale factor
$a$ of a (closed) Friedmann Universe and a homogeneous scalar field
$\phi$. The analogue of (3.11) is then a quantum mechanical
Schr\"odinger equation for a wave function $\psi(a,\phi)$ augmented
by two correction terms:
\be i\hbar\dot{\psi} =H_m\psi
    +\frac{G}{3\pi c^4 a}\left(H_m^2 +i\hbar H_0
    \left[-\frac{\partial H_m}{a\partial a} +H_m
    \right]\right)\psi. \lb{3.16} \ee
Here, $H_0\equiv \dot{a}/a$ denotes the Hubble parameter
of the Friedmann Universe, and $H_m$ is the Hamiltonian for
the scalar field. One recognises that the correction terms
only become relevant for small scales ($a\to 0$), i. e. in the
early Universe.

\subsection{Scalar field in de~Sitter space}

As a specific example we treat the case of a minimally
coupled scalar field, whose Hamiltonian density is given by
(2.25) with $U\equiv 0$. The case of non-minmally fields involves
extra subtleties and is treated elsewhere, see Kiefer (1993b).
It is convenient to write the three-metric in the following form:
\be h_{ab}=h^{1/3}\tilde{h_{ab}}, \lb{3.17} \ee
where $h$ is the determinant of the three-metric. (The "conformal part"
$\tilde{h_{ab}}$ thus has unit determinant.)
The Hamilton-Jacobi equation (2.29) then reads
\be -\frac{3\sqrt{h}}{16}\left(\frac{\delta S_0}
    {\delta\sqrt{h}}\right)^2 +\frac{\tilde{h_{ac}}
    \tilde{h_{bd}}}{2\sqrt{h}}\frac{\delta S_0}{\delta
    \tilde{h_{ab}}}\frac{\delta S_0}{\delta \tilde{h_{cd}}}
    -2\sqrt{h}(R-2\Lambda)=0. \lb{3.18} \ee
We want to look for a solution of this equation which, after
integrating (2.30), describes de~Sitter spacetime in a flat
foliation. This is easily achieved by setting $R=0$ in (3.18)
and looking for a solution of the form $S_0=S_0(\sqrt{h})$.
The desired solutions read
\be S_0=\pm 8\sqrt{\frac{\Lambda}{3}}\int\sqrt{h}d^3x
    \equiv \pm 8H_0\int\sqrt{h}d^3x, \lb{3.19} \ee
where, again, $H_0$ is the Hubble parameter (which is a constant
in this case).
According to (2.38), $S_0$ defines a local time parameter in
configuration space,
\[ \frac{\delta}{\delta\tau} =-\frac{3\sqrt{h}}{8}
   \frac{\delta S_0}{\delta\sqrt{h}}\frac{\delta}{\delta\sqrt{h}}
   =\sqrt{3h\Lambda}\frac{\delta}{\delta\sqrt{h}}, \]
from where the well kown expansion law for the scale factor
can be found,
\be \frac{\partial a^3}{\partial t} =
   \int d^3y\frac{\delta\sqrt{h}({\bf x})}{\delta\tau({\bf y})}
   =3H_0a^3\ \to \ a(t)= e^{H_0t}. \lb{3.20} \ee
Equation (2.36) for the prefactor can be immediately solved and
yields $\delta D/\delta\tau=0$, i. e. $D$ is constant on the
whole de~Sitter spacetime. The functional Schr\"odinger equation
(2.38) then reads in its integrated form (setting $\hbar=1$
in the following)
\be i\dot{\psi} =\int d^3x\left(-\frac{1}{2a^3}
    \frac{\delta^2}{\delta\phi^2} +\frac{a}{2}(\nabla\phi)^2
    +\frac{a^3}{2}m^2\phi^2\right)\psi. \lb{3.21} \ee
Since de~Sitter space is homogeneous, it is adequate to use
a momentum representation, i. e.
\be \phi({\bf x})=\int\frac{d^3k}{(2\pi)^3}\chi({\bf k})e^{i{\bf kx}}
    \equiv \int d\tilde{{\bf k}}\chi_k e^{i{\bf kx}}, \lb{3.22} \ee
and thus
\[ \frac{\delta}{\delta\phi({\bf x})}= \int d\tilde{k}e^{i{\bf kx}}
   \frac{\delta}{\delta\chi_k}\ ; \; \;
   \frac{\delta\chi_k}{\delta\chi_{k'}}
   =(2\pi)^3\delta^{(3)}({\bf k}+{\bf k'}). \]
We want to solve (3.21) by a Gaussian ansatz, i. e.
\be \psi=N(t)\exp\left(-\frac{1}{2}\int d\tilde{k}
    \Omega({\bf k},t)\chi_k\chi_{-k}\right). \lb{3.23} \ee
Gaussian states are used to describe generalised vacuum states
in the Schr\"odinger picture, see e. g. Jackiw (1988b). This special
form is of course tied to quadratic Hamiltonians like (2.25) with
$U=0$. According to (3.21), the Gaussian form (3.23) is
preserved in time.
Inserting this ansatz into (3.21) yields two equations for $N$
and for $\Omega$:
\bea i\frac{\dot{N}}{N} &=& \frac{1}{2a^3}\mbox{Tr}\Omega
     \equiv \frac{V}{2a^3}\int d\tilde{k}\ \Omega({\bf k},t)
     \lb{3.24} \\
     i\dot{\Omega} &=& \frac{\Omega^2}{a^3} -a^3(m^2 +
     \frac{k^2}{a^2}). \lb{3.25} \eea
Eq. (3.25) can be simplified by writing
\be \Omega=-ia^3\frac{\dot{y}}{y}, \lb{3.26} \ee
which leads to a linear second-order equation for $y$,
\be \ddot{y} +3H_0\dot{y}+ \left(m^2+\frac{k^2}{a^2}\right)
     y=0. \lb{3.27} \ee
It is useful to introduce a conformal time coordinate $\eta$,
\be dt=ad\eta \; \to \; a(\eta)=-\frac{1}{H_0\eta},
    \ \eta\in (0,-\infty),\lb {3.28} \ee
which yields
\be y''+2\frac{a'}{a}y' +(m^2a^2+k^2)y=0, \lb{3.29} \ee
where $'$ denotes a derivative with respect to $\eta$.
Inserting the evolution law (3.28) for the scale factor, this
equation can be explicitly solved in terms of Hankel functions,
see Guven, Lieberman, and Hill (1989). The question then arises
what boundary conditions one has to choose to select a particular
solution of (3.29). We will follow Guven {\em et al} and choose
the {\em Bunch-Davies vacuum}. This is de~Sitter invariant,
i. e. invariant under $SO(3,1)$, and reduces
to the Minkowski vacuum at early times, i. e. $\Omega$ tends to
$\sqrt{k^2+m^2}$ as $t\to-\infty$, where the metric is essentially
static and one can put $\dot{\Omega}=0$ and $a=1$ in (3.25).
The demand for de~Sitter invariance is the analogue of
the demand for Poincar\'{e} invariance in Minkowski space.
In $2+1$ dimensions it has been shown that the Bunch-Davies
vacuum is the {\em only} de~Sitter invariant vacuum state,
see Jackiw (1988b).

We are now interested to calculate the expectation value of the
Hamiltonian density with respect to (3.23). Using the solution
for $\Omega$ which corresponds to the Bunch-Davies vacuum, one
finds for $\langle\phi^2\rangle$,
\[ \langle\phi^2\rangle =\int\frac{d\tilde{k}}{2{\mbox Re}\Omega}
   \propto \frac{H_0^2}{4\pi^2}\int_0^{\infty}dyy, \]
i. e. a divergent result! The occurrence of ultraviolet divergencies
is of course not surprising, and we need a regularisation scheme
to extract physical predictions from (3.23). Guven {\em et al}
(1989) employ a dimensional regularisation scheme: They evaluate
all expressions in $d$ space dimensions, where they are finite,
and subtract terms which diverge in the limit $d\to 3$. After
some lengthy calculation they find a finite result for the expectation
value of the Hamiltonian which agrees with the result found
by using the effective action. In the massless case, for example,
one has (re-inserting $\hbar$ and $c$)
\be \langle {\cal H}_m\rangle =\frac{29\hbar H_0^4a^3}
     {960\pi^2 c^3}. \lb{3.30} \ee
This result is obtained on the level of the functional Schr\"odinger equation
(3.21), which is the level of quantum field theory on a classical
spacetime.\footnote{Note that $\langle{\cal H}_m\rangle
=a^3\langle T_{00}\rangle$, where $T_{00}$ is the $00$- component
of the energy-momentum tensor.}
 As we have seen in the last section, the next order of
the semiclassical approximation scheme yields corrections to the
Schr\"odinger equation. In the present case, Eq. (3.11) reads
\be i\hbar\frac{\delta\psi}{\delta\tau}
    ={\cal H}_m\psi -\frac{2\pi G}{c^4\sqrt{h}\Lambda}
    {\cal H}_m^2\psi -i\hbar\frac{2\pi G}{c^4\Lambda}
    \frac{\delta}{\delta\tau}\left(\frac{{\cal H}_m}{\sqrt
    {h}}\right)\psi. \lb{3.31} \ee
The first correction term in (3.31) leads to a shift in the above
expectation value, while the second term is a source of non-unitarity.
I want to focus first on the real term. It causes a shift
\be \langle {\cal H}_m\rangle \; \to \; \langle{\cal H}_m\rangle
    -\frac{2\pi G}{3c^2a^3H_0^2}\langle {\cal H}_m^2\rangle.
    \lb{3.32} \ee
The last term is of course utterly divergent and it is not clear
how to regularise it. On the covariant level it corresponds to
a term like $\langle T_{\mu\nu}^2(x)\rangle$, i. e. to a four-point
correlation function, whose regularisation in the presence of
gravitational fields has, to my knowledge, not been addressed.
Nevertheless, one can evaluate (3.32) in the present case, at
least in a heuristic sense. The reason is that the Bunch-Davies
vacuum state is an adiabatic vacuum state. The fluctuation
of such a state would be zero in quantum mechanics. I assume
that this holds also for the present field theoretic case, i. e.
that
\be \langle {\cal H}_m^2\rangle \approx \langle {\cal H}_m
    \rangle^2. \lb{3.33} \ee
The covariant version of (3.33) is frequently employed as a
criterion for semiclassical behaviour (Kuo and Ford, 1993). This
will be discussed in more detail in the next section.

One thus gets the following prediction for the shift in the
expectation value (3.30):
\be \langle {\cal H}_m\rangle \; \to \;\langle{\cal H}_m\rangle
    -\frac{841}{1382400\pi^3}\frac{G\hbar H_0^6a^3}{c^8}.
    \lb{3.34} \ee
Although found by using heuristic arguments, this is a definite
prediction of the Wheeler-DeWitt equation! That the correction term
is proportional to $GH_0^6$ is of course clear from dimensional
arguments. We remark already at this point that there will be
a second contribution to the energy shift which arises from
the modification of semiclassical time through the back reaction
of the matter fields.
Surprisingly, this modification will just lead to a change of
sign in the energy shift (3.34).
 The details will be presented in the next section.

What about the unitarity violating term?
Again, its regularisation is unclear, but one can make the following
rough estimate. It produces an imaginary contribution to the energy
density of the order of magnitude
\bea \vert \mbox {Im}\epsilon\vert &=& \frac{2\pi G\hbar}{3c^2H_0^2}
     \frac{\langle{\cal H}_m\rangle}{\tau Va^3}
     \nonumber\\
     & \approx & \frac{2\pi G\hbar }{c^2H_0V}
       \frac{\langle {\cal H}_m\rangle}{a^3}
       =\frac{29G\hbar^2 H_0^3}{480\pi Vc^5}. \lb{3.35} \eea
The presence of an energy density with an
imaginary part is of course
an indication of a possible instability of the system.
The associated time scale $t^*$ on which this "quantum gravitational
instability of de~Sitter space" may become relevant is given by
\be t^*= (2\mbox{Im}\epsilon V)^{-1}\hbar
    =\frac{240\pi c^5}{29G\hbar H_0^3}
    \sim \left(\frac{H_0^{-1}}{t_{Pl}}\right)^3,
    \lb{3.36} \ee
where $t_{Pl}$ is the Planck length. This time scale is thus only
relevant if the "horizon scale" $H_0^{-1}$ is of the order of
the Planck length.

\subsection{Evaporation of black holes}

We now turn to an example where the second, imaginary,
correction term in (3.11) becomes important. This will be the case
if the mass of a black hole approaches the Planck regime through
the emission of Hawking radiation (Kiefer, M\"uller, and Singh, 1993).

A black hole is produced by the collapse of matter, which we
can assume for simplicity to be in a pure state initially.
The presence of quantum fields on the black hole spacetime
leads to Hawking radiation, i. e. the black hole emits thermal
radiation and shrinks. What happens at the final
stage of this evaporation is an open issue. If the evaporation is
complete, only the Hawking radiation is left behind. Since thermal
radiation cannot, however, be described by a pure state, there
seems to be an evolution from pure states into mixed states.
This was taken as a hint that quantum gravity may
violate unitarity (Hawking, 1976) and that information is lost
at a fundamental level. There are, however, suggestions how
this "information loss paradox" can be resolved (a recent review
of this issue is Page (1993)). One possibility
is that the prescription through Hawking radiation is incomplete
and that the true time evolution is unitary. This would mean that
information must be encoded in the black hole radiation, for example by
stimulated emission or correlations. A second possibility is
that the black hole evaporation is not complete but ceases at a
mass of the order of the Planck mass, leaving behind a
Planck mass remnant. The main problem with this approach is
that only little energy is available to store a huge amount of
information.

The main drawback of all these approaches is that,
although the problem is a genuine problem of quantum gravity,
 only the semiclassical
theory has been considered. All scenarios put
forward so far thus only rest on speculations. To address these
issues properly, one should start from a specific approach
to quantum gravity such as the Wheeler-DeWitt equation (2.1).
While it is not yet clear how to apply this equation to the
black hole case one can at least apply the corrected
 Schr\"odinger equation, see (3.11), and hope that the
results obtained indicate some genuine features of quantum gravity.

As discussed above, the functional Schr\"odinger equation
(2.38) describes the level of quantum field theory on a
classical gravitational background, i. e. the level where the
Hawking effect is calculated, see, e. g., Freese, Hill, and Mueller
(1985) for a discussion in the functional Schr\"odinger picture.
We choose a slicing of the background spacetime, which is
obtained from $S_0$, such that $R=0$. How can the correction terms
in (3.11) then be evaluated? An important observation is
that one deals here with an {\em asymptotically flat} space,
in contrast to the compact case of the general discussion.
The Wheeler-DeWitt equation (2.1),
if integrated over space,
 is thus not correct as it stands
but has to be supplemented by a surface term from infinity.
This is nothing but the ADM mass, ${\cal M}$. One thus has to replace the
term $\sqrt{h}R$ in the denominators of (3.11) by
$-16\pi G{\cal M}/c^2$. This is already clear from dimensional
arguments, since $\sqrt{h}R$ has dimensions of a length and the only
length scale in the black hole case is the Schwarzschild radius.

One might wonder whether the functional Schr\"odinger equation
(2.38) can be recovered at all in this case, since the flat slicing
of Schwarzschild spacetime corresponds to a solution
$S_0=constant$ of the Hamilton-Jacobi equation, and the whole
spacetime would thus only be a point in the configuration
space. This would certainly be the case for a compact space,
but in the asymptotically flat case a distinguished time parameter
is associated with the Poincar\'e group at infinity. This
time parameter is the one that shows up in the
Schr\"odinger equation (2.38). It is also the time parameter
with respect to which the question of unitarity is discussed
in the case of black hole evaporation.

The ratio of the first correction term in (3.11) to the dominant
term, which is ${\cal H}_m$, is given by the ratio of the
energy of the scalar field to ${\cal M}c^2$, which is small even
if the mass of the hole approaches the Planck regime. This is not
the case for the second term! It contributes the following
term to the Hamiltonian:
\be \Delta H_m\equiv-\frac{4\pi iG\hbar}{c^4}
    \int d^3x{\cal H}_m\frac{\partial}{\partial t}
    \left(\frac{c^2}{16\pi G{\cal M}(t)}\right). \lb{3.37} \ee
We have here taken into account only the time derivative of
the ADM mass, since it is the change of the background geometry which
is the relevant contribution in the final stage of the evaporation
(the decrease of the ADM mass due to Hawking evaporation).
The time dependence of the mass of the black hole can be obtained
by using the expression for the Hawking temperature and
Stefan-Boltzmann's law, see e. g. DeWitt (1975). This leads
to
\be {\cal M}(t)= \left({\cal M}_0^3- \frac{a\hbar c^4}
    {G^2}t\right)^{1/3}, \lb{3.38} \ee
 where ${\cal M}_0$ is the initial mass of the hole, and $a$
 is a numerical factor that depends on the details of the model
 (the number of particle species considered, etc.) and whose precise
 value is not important for the present qualitative discussion.
Eq. (3.37) can of course only be used if one assumes that the black
hole does not settle to a mass which is much bigger than the Planck mass.

The term (3.37) becomes important if its of the same order of
magnitude than the Hamiltonian $H_m$ of the scalar field,
i. e. if
\be \frac{\hbar}{4M^2c^2}\frac{\partial{\cal M}}{\partial t}
    \approx \left(\frac{m_{Pl}}{{\cal M}}\right)^4
    \approx 1. \lb{3.39} \ee
The non-unitary term thus becomes important if the mass of the
evaporating black hole approaches the Planck mass.
In the above scenario, this happens
after a time
\be \bar{t} \approx\left(\frac{{\cal M}_0}{m_{Pl}}
    \right)^3t_{Pl}. \lb{3.40} \ee
After the mass of the hole has entered the Planck regime,
the semiclassical expansion breaks down and one would have
to deal with the full Wheeler-DeWitt equation. Nevertheless,
the correction term (3.37) may be a good approximation if,
for example, the black hole settles to a remnant of the
order of the Planck mass but such that the numerical value
of (3.37) is still small compared to $H_m$. It may also happen
that the final evaporation time is of the order of
${\cal M}_0^4$ (Carlitz and Willey, 1987) which would mean that
(3.37) is an excellent approximation over a long period of time.

A non-hermitean Hamiltonian does not transform a pure state
into a mixed state. It does, however, change the "degree of
purity" for a density matrix. In the present case its sign
is such that it {\em increases} the degree of purity,
if the mass of the hole decreases, see
Kiefer, M\"uller, and Singh (1993). Whether this can be interpreted
as a hint for the unitarity of quantum gravity is an open issue.

Although the above considerations are of a heuristic nature,
they nevertheless demonstrate that the correction terms to the
Schr\"odinger equation can be applied to concrete physical
examples.

\chapter{Decoherence and back reaction}

In the derivation of the Schr\"odinger equation from quantum
gravity, as it was presented in section~2, a crucial assumption was
that the gravitational part of the wave functional is of the
very special form $D^{-1}\exp(iMS_0/\hbar)$. The use of such
a special, complex, solution to the real Wheeler-DeWitt equation
(2.1) has been criticised in particular by Barbour (1993), who
regarded this choice as being equivalent to putting in the
desired result, namely the time-dependent
Schr\"odinger equation, by hand. The purpose of this section is to
demonstrate that the use of such a single WKB state can be
justified in a very natural, physical, way.

Consider, for example, a solution to (2.1) at order $M^0$
which is a superposition of the state (2.40) with its complex
conjugate,
\be \Psi=\frac{1}{D}\exp(iMS_0/\hbar)\chi[h_{ab},\phi]
    + \frac{1}{D}\exp(-iMS_0/\hbar)\chi^*[h_{ab},\phi]. \lb{4.1} \ee
Both $\chi$ and $\chi^*$ obey a time-dependent Schr\"odinger equation,
where time is defined respectively by $S_0$ and $-S_0$. The superposition
(4.1), however, does {\em not} describe a classical world, since
there are {\em interference terms} between both components of
(4.1). Note that the superposition (4.1) is analogous to the
superposition (2.6) in the derivation of the Born-Oppenheimer
approximation.

How can one get rid of these "unwanted" superpositions?
The basis idea is the fact that states like (4.1) are highly
correlated in the high-dimensional configuration space of
gravitational and non-gravitational degrees of freedom,
but that only very few of them are actually accessible to
a localised observer. Instead of the full state (4.1)
such an observer
 has thus to discuss the reduced density matrix, which is obtained
from (4.1) by integrating out the (huge number of) irrelevant
degrees of freedom. States of the accessible part, which
are then correlated with orthogonal states of the "rest of the
world" do not exhibit any interference terms locally. Information
about the corresponding phase relations has been {\em delocalised}
in the full configuration space. This mechanism is referred to as
{\em decoherence}. In the derivation of the Born-Oppenheimer
approximation (2.8), decoherence lies at the heart of the
neglection of interference terms.

In the following I will first briefly discuss two
examples from quantum mechanics and then show how decoherence
can justify the use of only one of the components in (4.1).
For a general discussion on decoherence I refer to the
extensive literature on this subject - see, for example,
Zeh (1992, 1993a) or Zurek (1991), and the references therein.

\section{Decoherence in quantum mechanics: Two examples}

The first example is the {\em localisation} of macroscopic
objects through interaction with their natural environment,
for example through scattering by photons or air molecules
(Joos and Zeh, 1985). Be $\vert{\bf x}\rangle$ the centre of mass
position eigenstate of the
macroscopic body, and $\vert\chi\rangle$ an initial
state of the environment such that the total state at, say,
$t=0$ is a product of these states. The interaction be such
that this state evolves according to
\be \vert{\bf x}\rangle \vert\chi\rangle \; \stackrel{t}{\to} \;
    \vert{\bf x}\rangle S_{{\bf x}}\vert\chi\rangle,
    \lb{4.2} \ee
where $S_{{\bf x}}$ denotes the scattering matrix. The state of
the environment is thus correlated with the state of the body.
If one starts now from an entangled, nonclassical, state
for the body, it will envolve, according to the superposition
principle, into an entangled state of the body with the
environment, i. e.
\be \left(\int d^3x \varphi({\bf x})\vert{\bf x}\rangle
    \right)\vert\chi\rangle \; \stackrel{t}{\to}
    \; \int d^3x \varphi({\bf x})\vert{\bf x}\rangle
    S_{{\bf x}}\vert\chi\rangle. \lb{4.3} \ee
There is no sign of any localisation for the body in this state!
Since, however, the huge number of degrees of freedom of the
environment is inaccessible to a local observer, the relevant
object is the density matrix of the position of the scattering
centre after scattering:
\bea \rho({\bf x}, {\bf x'}) &=&
     \varphi({\bf x})\varphi^*({\bf x'})
     \langle\chi\vert S_{{\bf x'}}^{\dagger}S_{{\bf x}}
     \vert\chi\rangle \nonumber\\
     & \approx & \varphi({\bf x})\varphi^*({\bf x'})
      \exp\left(-\Lambda t({\bf x}-{\bf x'})^2\right). \lb{4.4} \eea
The Gaussian factor in (4.4) leads to the suppression of
interference effects between different positions of the
scattered object ($\Lambda$ contains the details of the
scattering process) -- the object becomes localised,
if the mass of the body is sufficiently large, and
wave packets are prevented from spreading.
The density matrix in (4.4) obeys a non-unitary master equation
instead of a unitary von Neumann equation.
 As shown by
Joos and Zeh (1985), even the scattering by the photons of the
microwave background leads to the localisation of a dust grain
in interstellar space.

Two important aspects of decoherence should be stressed
(Zeh, 1993b). The first aspect follows from the very formalism of
decoherence and is independent of any interpretation. It states
that the phase relations of an entangled state like (4.3)
can never be seen by a local observer according to the
{\em probabilistic rules} used in conventional quantum theory.
This does {\em not} mean, however, that only one component
of this state survives -- all components are still "there."
The second aspect addresses the issue of why only one
component is "observed." This involves the interpretation of the
quantum formalism. Basically, there are two options:
Either one has to assume that the decohered components are
still co-existing, but are dynamically independent from each
other (this dynamical independence
 is what decoherence achieves) or one has to invoke
explicit collapse models (and thereby has to {\em change}
conventional quantum theory) to single out a specific component,
which is then the "observed one." For all practical purposes,
this is equivalent to the branching into dynamically independent
components in the first interpretation, where decoherence
defines an ensemble of wave functions, the "preferred basis"
(Zurek, 1993).

The second example is taken from nuclear physics, where deformed
nuclei exhibiting
a definite orientation
 can be found even if the total quantum state is an angular
momentum eigenstate (Zeh 1967, 1993b). Consider an $N$-nucleon
system which is described by a {\em stationary} Schr\"odinger
equation. An {\em approximate} solution,
as it is found through an appropriate variation principle,
 is given by the antisymmetrised
produced, $\phi$, of one-particle nucleon wave functions
(Slater determinants). Since a superposition of different such
$\phi$ is {\em not} a determinant again (due to the
 nonlinearity of the ansatz),
these solutions need not be eigenstates of an operator like
angular momentum which commutes with the Hamiltonian. They may
thus describe a deformed nucleus, which has a definite orientation.
Performing a superposition of different "deformed solutions,"
\be \Psi=\int d\Omega f(\Omega)U(\Omega)\phi, \lb{4.5} \ee
where $U(\Omega)$ denotes a unitary representation of the rotation
group and $\Omega$ may stand, e. g., for the Euler angles, one
obtains, of course, again an approximate solution to the stationary
Schr\"odinger equation. The amplitude $f(\Omega)$ may be chosen such
that $\Psi$ is an approximate eigenstate of angular momentum.

If many nucleons are present in the nucleus, a certain nucleon is
correlated with a corresponding deformed potential generated by
the remaining nucleons in each component of the superposition
(4.5) ("strong coupling"). The matrix element between different
orientations reads\footnote{We assume here for simplicity that
$\phi=\prod\phi_i$ (no antisymmetrisation).}
\bea \langle U(\Omega)\phi\vert U(\Omega')\phi\rangle
     &=& \prod_{i=1}^N \langle U(\Omega)\phi_i\vert
     U(\Omega')\phi_i\rangle \nonumber\\
     & \propto& \exp\left(-\lambda (\Omega-\Omega')^2\right).
     \lb{4.6} \eea
For large $N$ this is a narrow Gaussian peaked around
$\Omega=\Omega'$ -- the different orientations in (4.5) have
decohered from one another.
This is an example of a spontaneous symmetry breaking.\footnote{Other
important examples are chiral molecules and the Higgs vacuum,
see Zeh (1992).}
 As can be seen from the existence
of a rotational spectrum, these different orientations even
seem to rotate slowly, in spite of the stationary nature of the
total state! This is the analogue to the recovery of an
approximate time-dependent Schr\"odinger equation for intrinsic degrees
of freedom from the stationary Wheeler-DeWitt equation (2.1).
Since $\langle U(\Omega')\phi\vert U(\Omega)\phi\rangle
\approx \delta(\Omega-\Omega')$, one has $\langle \phi(\Omega)
\vert\psi\rangle\approx f(\Omega)$ and can thus assign an approximate
wave function $f(\Omega)$ to the rotational degrees of freedom,
describing a fictitious "rigid top."

If the nucleus contained an intrinsic quantum observer, such
an observer would see the nucleus with a definite orientation.
In the case of quantum cosmology, where the Wheeler-DeWitt equation
(2.1) is applied to the whole Universe, all observers are
intrinsic observers, and decoherence is the reason why they
observe a classical world. This is discussed in the next subsection.

\section{Decoherence of different WKB branches in quantum
            gravity}

The analogue of the rotational symmetry of the nuclear Hamiltonian
is here the (discrete) symmetry of the Wheeler-DeWitt Hamiltonian
in (2.1) under complex conjugation. The semiclassical expansion
presented in the previous sections shows that, to order $M^0$,
the functionals $\chi[h_{ab},\phi]$ of the non-gravitational
fields obey an approximate time-dependent Schr\"odinger equation,
which is a {\em complex} equation! Consequently,
$\chi$ is in general complex and $\chi$ and $\chi^*$
thus couple drastically different to the components $\exp(iMS_0
/\hbar)$ and $\exp(-iMS_0/\hbar)$. Since a huge number
of degrees of freedom is involved, this leads to decoherence
between the two components in (4.1), if these degrees of freedom
are integrated out to get the reduced density matrix
\be \rho[h_{ab},h_{ab}'] \equiv
 \mbox{Tr}_{\phi}\Psi[h_{ab},\phi]\Psi^*[h_{ab}',\phi]
 \lb{4.7} \ee
for the gravitational field.
This general expectation is verified in concrete models
(see Kiefer (1992c) for a review). Typically, global degrees
of freedom like the scale factor $a$ of a Friedmann Universe are
chosen to represent the semiclassical gravitational degrees
of freedom, while "small fluctuations" like density perturbations
or gravitational waves play the role of the irrelevant degrees
of freedom, which are integrated out. The functional $\chi$ is
written in these examples as
\be \chi=\prod_{n=1}^N \chi_n(a;x_n), \lb{4.8} \ee
where the $x_n$ may be, for example, coefficients
in the expansion of the scalar field $\phi$ into its harmonics.
The result is that $\rho$ becomes a narrow Gaussian in
$a-a'$ (this is the analogue of the quantum mechanical
example on localisation and may here
be described as a "localisation within
one WKB component" or
"measurement of the three-geometry") and contains in addition a large
suppression factor for the interference term between the
WKB components in (4.1) (this is the analogue of the deformed
nucleus or the chiral molecule). There are, however, situations
where $\chi$ is approximately real and the two components in
(4.1) interfere. This happens, for example, in regions which would
correspond in the classical theory to the turning point of
a recollapsing universe (Kiefer, 1992b), where
$\partial S_0/\partial a$ becomes small.
Decoherence thus justifies the selection of a definite
WKB component in the semiclassical approximation as well as
the localisation within one component. This is, however,
not yet sufficient to justify the notion of a classical
space{\em time}. It requires in addition the persistence
of decoherence along a whole WKB trajectory and is related to the
issue of back reaction which will be discussed in the next
subsection.

I want to remark finally that it is of course not necessary
to deal only with the whole Universe in quantum gravity.
A quantum black hole, for example, may be viewed from outside
and is thus a more direct analogue of the deformed nucleus.

\section{Back reaction}

In this section I want to address the question to what extent the
presence of the non-gravitational matter fields lead to a
modification of the Hamilton-Jacobi equation (2.29) at order
$M^{-1}$, i. e. the question of back reaction. In particular, I want
to discuss the conditions under which this back reaction is given by
the {\em expectation value} of the Hamiltonian (2.25) with respect to the
state $\chi$ satisfying the approximate Schr\"odinger equation.
This is equivalent to study the range of validity of the
semiclassical Einstein equations
\be G_{\mu\nu}=\frac{8\pi G}{c^4}\langle T_{\mu\nu}\rangle.
    \lb{4.9} \ee
At order $M$, the value of the geometrodynamical momentum when applied
to the WKB state of the gravitational field is given by the familiar
expression (2.30). To find out how this expression may be changed
at the next order, we first decompose $\chi$ into its absolute
value and its phase,
\be \chi\equiv R\exp(i\theta/\hbar), \lb{4.10} \ee
and then rewrite the Hamilton-Jacobi equation (2.29) as follows:
\be \frac{1}{2M}G_{ab}\left(M\frac{\delta S_0}{\delta h_a}
    +\frac{\delta\theta}{\delta h_a}\right)
    \left(M\frac{\delta S_0}{\delta h_b}+\frac{\delta\theta}
    {\delta h_b}\right) +MV -\frac{\delta\theta}{\delta\tau}
    +{\cal O}(M^{-1})=0. \lb{4.11} \ee
Performing the expectation value of this equation with respect
to $\chi$ and introducing the quantity
\be \Pi^a \equiv M\frac{\delta S_0}{\delta h_a} +
   \langle\chi\vert\frac{\delta\theta}{\delta h_a}\chi\rangle,
   \lb{4.12} \ee
this can be written as
\be \frac{1}{2M}G_{ab}\Pi^a\Pi^b +MV
   -\langle\chi\vert\frac{\delta\theta}{\delta\tau}\chi\rangle
   +{\cal O}(M^{-1})=0. \lb{4.13} \ee
The crucial assumption made at this stage is that
$\Pi^a$ is the geometrodynamical momentum at this order of
approximation. We will specify later the conditions under which
this assumption is valid. Taking this for granted, Eq. (4.13)
is the desired "back reaction corrected" Hamilton-Jacobi equation.
It can be conveniently rewritten by noting that
\be \langle\chi\vert \frac{\delta\theta}{\delta\tau}
    \chi\rangle =-\langle\chi\vert {\cal H}_m\chi\rangle,
    \lb{4.14} \ee
which follows immediately from the fact that $\theta$ is the phase
of a solution to the time-dependent Schr\"odinger equation (2.38).
One can thus write (4.13) as
\be \frac{1}{2M}G_{ab}\Pi^a\Pi^b +MV
   +\langle\chi\vert{\cal H}_m\chi\rangle
   +{\cal O}(M^{-1})=0. \lb{4.15} \ee
 Making use of (2.30) and (2.38) one can write
 \be \langle\chi\vert{\cal H}_m\chi\rangle
     =-\frac{\hbar}{M}G_{ab}\Pi^aA^b, \lb{4.16} \ee
 where we have introduced in analogy to (2.9) a
 "super Berry connection" (see
 Balbinot, Barletta, and Venturi, 1990; Datta, 1993; Kiefer, 1993a)
 \be A^b= -i\langle\chi\vert\frac{\delta\chi}{\delta h_b}\rangle
    \lb{4.17} \ee
 which is defined on the space of three-metrics.

An equivalent way to write (4.13) is then, up to terms of the order
$M^{-1}$,
\bea & & \left\{ \frac{1}{2M} G_{ab}\left(-i\hbar\frac{\delta}{\delta h_a}
    -\hbar A^a\right)\left(-i\hbar\frac{\delta}{\delta h_b}
    -\hbar A^b\right)  +MV\right. \nonumber\\ & & \;
   \left. -\frac{\hbar^2}{2M}G_{ab}A^aA^b +{\cal O}(M^{-1})
    \right\}\psi_G=0, \lb{4.18} \eea
 where $\psi_G$ is an effective gravitational wave function
 at this order of approximation. I emphasize that this equation is very
 similar to the equation (2.8) for the nuclear wave function in the
 Born-Oppenheimer approximation. The term quadratic in the connection,
 which is already of order $M^{-1}$,
 has only been retained in (4.18) because of this comparison.

An important consequence of the back reaction is that the notion
of WKB time is modified since an additional term proportional
to $M^{-1}$ arises from (4.12). This term is
 thus of the same order as the corrections (3.11) to
the Schr\"odinger equation. Using the definition (4.12), one can rewrite
the time derivative introduced in (2.38) as follows:
\bea i\hbar\frac{\delta\psi}{\delta\tau} & \equiv &
     i\hbar G_{ab}\frac{\delta S_0}{\delta h_a}
     \frac{\delta\psi}{\delta h_b} =
     i\hbar G_{ab}\left(\frac{\Pi^a}{M}-\frac{1}{M}
     \langle\chi\vert\frac{\delta\theta}{\delta h_a}\chi\rangle
     \right)\frac{\delta\psi}{\delta h_b} \nonumber\\
     &\equiv & i\hbar\frac{\delta\psi}{\delta\tilde{\tau}}
     -\frac{i\hbar}{M} G_{ab}\langle\chi\vert
     \frac{\delta\theta}{\delta h_a}\chi\rangle\frac{\delta\psi}
     {\delta h_b}. \lb{4.19} \eea
At order $M^{-1}$ the "back reaction corrected" WKB time
$\tilde{\tau}$ has to be used instead of the old $\tau$. To find the induced
contribution of this change in the definition of time to the corrected
Schr\"odinger equation (3.11), we first decompose the vector field
$G_{ab}\delta\psi/\delta h_a$ as in (3.9) into its components
tangential and orthogonal to the flow lines of the vector field
$G_{ab}\delta S_0/\delta h_a$ and assume, again, that the orthogonal
component is much smaller than the tangential component. Thus,
\be G_{ab}\frac{\delta\psi}{\delta h_a}
    =\frac{i}{2\hbar V}{\cal H}_m\psi G_{ab}\frac{\delta S_0}
    {\delta h_a} +{\cal O}(M^{-1}). \lb{4.20} \ee
With the help of (4.14) one can then write (4.19) in the form
\be i\hbar\frac{\delta\psi}{\delta\tau} =
    i\hbar\frac{\delta\psi}{\delta\tilde{\tau}}-
    \frac{1}{2MV}\langle\chi\vert{\cal H}_m\chi\rangle
    {\cal H}_m\psi, \lb{4.21} \ee
so that the corrected Schr\"odinger equation (3.11) can be written
as an equation
with respect to the new time $\tilde{\tau}$:
\bea i\hbar\frac{\delta\psi}{\delta\tilde{\tau}}
     &=& {\cal H}_m\psi +\frac{4\pi G}{c^4\sqrt{h}
       (R-2\Lambda)}({\cal H}_m^2-2 \langle {\cal H}_m\rangle
       {\cal H}_m)\psi \nonumber\\
     & & \; +i\hbar\frac{4\pi G}{c^4}\frac{\delta}{\delta\tau}
     \left(\frac{{\cal H}_m}{\sqrt{h}(R-2\Lambda)}\right)
     \psi. \lb{4.22} \eea
The modification in the definition of time thus leads to an
additional contribution to the shift of energies induced by quantum
gravity. In our example of the scalar field in de~Sitter space
(section 3.2.2) this additional term leads to a {\em change
of sign} in the shift of the expectation value for the
Hamiltonian density. Instead of (3.34) one has
\be \langle{\cal H}_m\rangle \; \to \;
    \langle{\cal H}_m\rangle +\frac{841}{1382400\pi^3}
    \frac{G\hbar H_0^6a^3}{c^8}. \lb{4.23} \ee
{}From the corrected Hamilton-Jacobi equation (4.15) one finds in
this example, using (3.30), an effective {\em increase} of the Hubble
parameter
\be H_0^2 \; \to \; H_0^2\left(1+\frac{29}{360\pi}
    \frac{GH_0^2\hbar}{c^5}\right). \lb{4.24} \ee
The increase in $H_0$ is thus essentially determined by the
square of the ratio of the Planck length to the horizon length.
If our present Universe were in a de~Sitter phase this
 increase
in $H_0$ would be of the order of $10^{-120}$.

Analogus to the Hamilton-Jacobi equation one gets a correction
to the momentum constraints at this order of approximation, i. e.
\be D_a\Pi^{ab}=\langle\chi\vert T^b_0\chi\rangle . \lb{4.25} \ee
One has thus recovered the semiclassical Einstein equations (4.9)
at order $M^{-1}$.

What are the exact conditions under which the semiclassical
Hamilton-Jacobi equation is valid? The first, obvious, condition
is that the expectation value of ${\cal H}$ with respect to the
state satisfying the approximate Schr\"odinger equation be {\em small}
compared to the other terms in (4.15). This came out in a natural
way through the expansion scheme with respect to the Planck mass.
The second condition, as has been discussed in section~4.1,
is the smallness of interference terms between different WKB
components of the total state. This is the analogy of the
smallness of interferences between the different components of (2.6)
and is a necessary condition for the validity of a
Born-Oppenheimer approximation. As shown in section~4.1, these
various components decohere in realistic cases and thereby justify
the use of a single WKB component.

 The third, and most subtle,
condition is the possibility to replace the geometrodynamical momentum
at order $M^{-1}$ by the expression (4.12).
This can be done if there is a strong correlation in the
gravitational sector between the momentum and the expression on
the right-hand side of (4.12). There is no simple general
condition for this to be the case. The usual approach is
to employ the {\em Wigner function} for the reduced density matrix
(4.7) obtained by integrating out non-gravitational degrees
of freedom within a {\em single} WKB component. For a given
density matrix in the position representation $\rho(q,q')$,
the Wigner function is defined as
\be F_W(\bar{q},p)=\int d\Delta e^{-2ip\Delta}
    \rho(\bar{q}+\Delta,\bar{q}-\Delta), \lb{4.26} \ee
where $\bar{q}=(q+q')/2$ and $\Delta=(q-q')/2$. The Wigner function
is the closest one can get to a classical phase space distribution;
because of the quantum uncertainty, the Wigner function is in general
not positive definite. The Wigner function has been used in
quantum gravity by several authors including Halliwell (1987),
Padmanabhan and Singh (1990), Habib and
Laflamme (1990), and Paz and Sinha (1991). To evaluate (4.26) one
has to first calculate the reduced density matrix by using a
definite solution of the Schr\"odinger equation satisfied by $\chi$.
In most examples discussed so far a Gaussian solution such as
(3.23) has been chosen. The Wigner function (4.26) can then
be evaluated and is in these examples typically of the form
\bea F_W[h_a,\Pi^a] &\approx & D^{-2}\sqrt{\frac{\pi}
    {\sigma^2}}\exp\left[-\frac{1}{\sigma^2}G_{ab}
    \left(\Pi^a-M\frac{\delta S_0}{\delta h_a}
    -\langle\chi\vert\frac{\delta\theta}{\delta h_a}\chi\rangle\right)
    \right. \nonumber\\ & & \; \left.
    \left(\Pi^b-M\frac{\delta S_0}{\delta h_b}
    -\langle\chi\vert\frac{\delta\theta}{\delta h_b}\chi\rangle\right)
   \right], \lb{4.27} \eea
  where
\be \sigma^2 =\mbox {Tr}\frac{\vert\delta\Omega/\delta\tau
     \vert^2}{4\mbox{Re}\Omega^2} \lb{4.28} \ee
turns out to be the inverse of the coherence width
associated with the reduced density matrix
itself. ($\Omega$ is the covariance of the Gaussian, see
(3.23).)
 The conditions for the validity of the semiclassical
Hamilton-Jacobi equation arising from these considerations are
thus a) that the width (4.28) of the Gaussian (4.27) is much smaller
than the peak value of the Gaussian ("strong correlation
condition") and b) that the inverse width is much smaller
than the components of the three-metric which are "measured" by
the huge number
of irrelevant degrees of freedom
("strong decoherence condition"). Typically, these conditions
are fulfilled if the {\em fluctuation of the Hamiltonian density
is small} in the state $\chi$ satisfying the approximate Schr\"odinger
equation, i. e. if (3.33) holds. Under this condition, therefore,
one would expect the expressions derived within this framework
to hold very accurately. It is, however, still unclear what are
the {\em most general conditions} which are compatible with
the back reaction-corrected Hamilton-Jacobi equation (4.15).

\chapter{Discussion}
It is my purpose in this contribution to present, as I hope
convincing, arguments that it is possible to extract physical
insight from semiclassical gravity even if the underlying exact
theory is not yet available. The formal framework is based on the
canonical approach to quantum gravity and its central
 equation (2.1). Although the fundamental Hamiltonian
may not be exactly of the form (2.1), general principles like
the reparametrisation invariance of the classical theory suggest
that a constraint equation of the form $H\Psi=0$ will still play
a central role. Semiclassical considerations may thus well be,
at least to a large extent, independent of the concrete form of the
fundamental equation. Moreover, the necessary requirement that
the, well-tested, Schr\"odinger equation can be recovered as an
approximate equation puts an important restriction on the fundamental
theory. Corrections to the Schr\"odinger equation, on the other side,
may turn out to be different from one theory to the other and
thus allow the possibility to distinguish between alternative
approaches. I emphasise that such correction terms lead to
definite physical predictions such as the energy shift in (4.23)
which can, at least in principle, be checked and thus confirmed
or falsified.

I want to finally give a brief outlook on other important topics
in semiclassical gravity which have not been addressed in this
contribution. The first topic has to do with the relevance
of semiclassical considerations for the problem of time in
quantum gravity (Isham, 1992; Kucha\v{r}, 1992). Briefly
speaking, the problem of time in quantum gravity is concerned
with the "timeless nature" of the Wheeler-DeWitt equation (2.1)
(by which one means the absence of the classical
time parameter $t$) and the question whether a physical
concept of time can be introduced at the most fundamental level
or only in the semiclassical regime. I have presented an
 approximation scheme
where the Schr\"odinger equation together
with its classical time parameter emerges in an appropriate limit.
The expansion scheme is only valid if the higher order terms are
small compared to the dominant one. The semiclassical Einstein
equation (4.9), in particular, is only valid if the right-hand
side is treated as a small perturbation. A nonperturbative treatment would
in general not lead to sensible results. This argument also
applies to the corrected Schr\"odinger equation (3.11). One would thus
not expect that this semiclassical notion of time can be
extrapolated to the full theory. On the contrary, the fact that
the Schr\"odinger equation and with it all the concepts of ordinary
quantum theory like Hilbert space and unitarity
 are only recovered
approximately (recall (3.14)) leads me to the suspicion that these concepts
do {\em not} play any fundamental role in quantum gravity.
As mentioned before, the presence of a Hilbert space structure
on the level of the exact theory may be in conflict with
results from semiclassical expansions, see Louko (1993).
Interpretations of quantum gravity
which only work in the semiclassical regime, like the one by Vilenkin (1989),
do in my opinion not really solve the problem of time and may even
present an obstacle on the way to finding the correct full theory.

There is one attempt to extrapolate the semiclassical concept
of time to full quantum gravity which has gained some attention,
see Padmanabhan (1990), Greensite (1990), and Squires (1991). The basic idea
there
is to extract time from the phase of the {\em full} wave functional.
In the semiclassical regime it agrees, of course, with the WKB time
in the approximate Schr\"odinger equation, but it can be drastically
different in other cases. This "phase time" parametrises, at least
locally, a flow in the configuration space. This flow does not,
in general, have anything to do with the background spacetime which was
reconstructed from a given solution to the Hamilton-Jacobi equation
(2.29). The main advantage of this approach seems to be that
an exact Ehrenfest equation can be written down for the expectation
value of an operator depending on the variables which are
comoving with respect to the "phase time flow." The expectation
value is performed with respect to the Schr\"odinger inner product
of the full wave functional and the comoving variables.
There are, however, some problems with this
approach. Firstly, the procedure does not work for real
wave functionals which can therefore not be interpreted in this
framework. Secondly, the concept of phase time depends on
the given solution for the Wheeler-DeWitt equation and may thus
seem inappropriate as a fundamental concept. Thirdly, phase time
does not exist globally and may also lead to the "spacetime
problem," i. e. the dependence of its definition upon the
choice of the original spacetime foliation in the canonical
approach (Kucha\v{r}, 1992). I emphasise again that this problem
did not occur in the semiclassical approximation. Finally,
the Ehrenfest condition with respect to the momentum variables
does only hold if the phase of the wave functional varies rapidly
along the flow lines which it defines.
In the special case, however, where the semiclassical expansion
scheme converges, it seems that the sum over all orders leads
to a modification in the original WKB time which just yields
the above phase time. This is spelled out elsewhere (Kiefer, 1993c).

Can anything be said about the concept of time on the
fundamental level of (2.1) itself? The answer is yes, since the
structure of the DeWitt metric in front of the kinetic term
distinguishes a certain variable in configuration space from
the remaining ones by its minus sign. At each space point,
(2.1) is thus of the hyperbolic type. But only in the case where
{\em one} global minus sign survives after the
diffeomorphism group has been factored out could one
say that (2.1) itself is hyperbolic. This is, however,
not the case in the full configuration space, but only
in special domains (Giulini, 1993). One important example is
the neighbourhood of Friedmann-type three-geometries
for constant lapse. In this
case there exists a global intrinsic timelike variable,
the scale factor of the Friedmann space, with respect to which
a well-defined initial value problem can be posed for the
wave functional (Zeh 1988, 1992a). There is, however, no unitary
evolution with respect to this intrinsic time.
This fact leads to an interesting tension between intrinsic time
and WKB time which has drastic
consequences in the case of a recollapsing
Friedmann Universe (Zeh, 1992; Kiefer, 1993d): The semiclassical
approximation can {\em not} be valid along the whole region
of the corresponding classical trajectory and genuine quantum effects
of gravity become important far away from the Planck scale.

I want to finally make a brief comparison of the
semiclassical approximation to quantum gravity in the canonical
approach with the covariant effective action approach
("background field method"). Formally, the level of the
approximate functional Schr\"odinger equation corresponds to
the level of the one-loop effective action. Taking into account
the correction terms of (3.11) is analogous to
the two-loop level of the effective action, although this
has not yet been explicitly demonstrated. There is, however,
a conceptual difference. The effective action approach starts
from a {\em given} background spacetime which in the canonical
framework was only recovered
under appropriate circumstances such as the decoherence between
different WKB components. The background field method therefore
can be applied only under these circumstances.

\vspace{5mm}

\begin{center}
{\bf Acknowledgement}
\end{center}
I thank the organisers for inviting me to this workshop and
the participants for many interesting discussions. I have attempted
to address their questions and comments in my contribution to these
Proceedings.

\vspace{1cm}

\begin{center}
{\bf References}
\end{center}

\begin{description}
\item[A. Ashtekar] (1991): {\em Lectures on non-perturbative canonical gravity}
                           (World Scientific, Singapore).
\item[A. Ashtekar] (1988): {\em New perspectives in canonical
                           gravity} (Bibliopolis, Napoli).
\item[R. Balbinot, A. Barletta, and G. Venturi] (1990):
   Matter, quantum gravity, and adiabatic phase. Phys. Rev. D
   {\bf 41}, 1848.
\item[T. Banks] (1985): TCP, Quantum Gravity, the Cosmological Constant,
                     and all that \ldots . Nucl. Phys. {\bf B249}, 332.
\item[J. B. Barbour] (1992): Private communication.
\item[J. B. Barbour] (1993): Time and complex numbers in canonical
                      quantum gravity. Phys. Rev. D {\bf 47}, 5422.
\item[E. P. Belasco and H. C. Ohanian] (1969): Initial Conditions
 in General Relativity: Lapse and Shift Formulation.
 Journ. Math. Phys. {\bf 10}, 1503.
\item[O. Bertolami] (1991): Nonlinear corrections to quantum mechanics
   from quantum gravity. Phys. Lett. {\bf A154}, 225.
\item[R. Brout and G. Venturi] (1989): Time in semiclassical
      gravity. Phys. Rev. D {\bf 39}, 2436.
\item[R. D. Carlitz and R. D. Willey] (1987): Lifetime of a black hole.
 Phys. Rev. D {\bf 36}, 2336.
\item[D. Cosandey] (1993): Localized quantum mechanics in a
 semiclassical universe. University of Bern preprint.
\item[T. Damour and J. H. Taylor] (1991): On the orbital period change of
        the binary pulsar PSR 1913+16. Astrophys. Journal {\bf 366}, 501.
\item[D. P. Datta] (1993): Semiclassical backreaction and
     Berry's phase. Mod. Phys. Lett. A {\bf 8}, 191 (1993).
\item[B. S. DeWitt] (1967): Quantum Theory of Gravity I.
             The Canonical Theory. Phys. Rev. {\bf 160}, 1113.
\item[B. S. DeWitt] (1970): ``Spacetime as a sheaf of geodesics
  in superspace", in {\em Relativity}, ed. by M. Carmeli, S. Fickler,
   and L. Witten (Plenum, New York).
\item[B. S. DeWitt] (1975): Quantum field theory in
  curved spacetime. Phys. Rep. {\bf 19}, 295.
\item[K. Freese, C. T. Hill, and M. Mueller] (1985): Covariant functional
 Schr"\-ding\-er formalism and application to the Hawking effect.
 Nucl. Phys. {\bf B255}, 693.
\item[U. H. Gerlach] (1969): Derivation of the Ten Einstein Field
  Equations from the Semiclassical Approximation to
  Quantum Geometrodynamics. Phys. Rev. {\bf 177}, 1929.
\item[D. Giulini] (1993): ``What is the geometry of superspace?",
   in {\em from Newton's bucket to quantum gravity},
   ed. by J. B. Barbour and H. Pfister (Birkh\"auser, Boston).
\item[J. Greensite] (1990): Time and Probability in
 Quantum Cosmology. Nucl. Phys. {\bf B342}, 409.
\item[J. Guven, B. Lieberman, and C. T. Hill] (1989): Schr\"odinger-picture
      field theory in Robertson-Walker flat spacetimes.
      Phys. Rev. D {\bf 39}, 438.
\item[S. Habib and R. Laflamme] (1990): Wigner function and decoherence
  in quantum cosmology. Phys. Rev. D {\bf 42}, 4056.
\item[J. J. Halliwell] (1987): Correlations in the wave function
   of the Universe. Phys. Rev. D {\bf 36}, 3626.
\item[J. J. Halliwell and S. W. Hawking] (1985): Origin of structure
 in the Universe. Phys. Rev. D {\bf 31}, 1777.
\item[J. B. Hartle] (1986): ``Prediction in Quantum Cosmology", in
 {\em Gravitation in Astrophysics}, ed. by B. Carter and
  J. B. Hartle (Plenum Press, New York).
\item[S. W. Hawking] (1976): Breakdown of predictability in
 gravitational collapse. Phys. Rev. D {\bf 14}, 2460.
\item[W. Heisenberg and H. Euler] (1936): Folgerungen aus der Diracschen
 Theorie des Positrons. Z. Phys. {\bf 98}, 714.
\item[C. J. Isham] (1992): ``Canonical Quantum Gravity and the Problem
 of Time", in
  {\em Integrable Systems, Quantum Groups, and
  Quantum Field Theories} (Kluwer Academic Publishers, London).
\item[R. Jackiw] (1988a): Berry's phase -- Topological Ideas
  from Atomic, Molecular and Optical Physics. Comments At. Mol. Phys.
   {\bf 21}, 71.
\item[R. Jackiw] (1988b): ``Analysis on infinite-dimensional
   manifolds -- Schr\"odinger representation for quantized
   fields", {\em Field Theory and Particle Physics}, ed. by
   O. Eboli, M. Gomes, and A. Santano (World Scientific,
   Singapore).
\item[E. Joos and H. D. Zeh] (1985): The Emergence of Classical
  Properties through Interaction with the Environment.
  Z. Phys. {\bf B59}, 223.
\item[C. Kiefer] (1987): Continuous measurement of mini-superspace variables
              by higher multipoles. Class. Quantum Grav. {\bf 4}, 1369.
\item[C. Kiefer] (1992a): Functional Schr\"odinger equation
  for scalar QED. Phys. Rev. D {\bf 45}, 2044.
\item[C. Kiefer] (1992b): Decoherence in quantum electrodynamics and
              quantum gravity. Phys. Rev. D {\bf 46}, 1658.
\item[C. Kiefer] (1992c): ``Decoherence in quantum cosmology", in
 {\em Proceedings of the 10th Seminar on Relativistic
 Astrophysics and Gravitation}, ed. by S. Gottl\"ober, J. P. M\"ucket,
 and V. M\"uller
 (World Scientific, Singapore).
\item[C. Kiefer] (1993a): Topology, decoherence, and semiclassical gravity.
                         Phys. Rev. D {\bf 47}, 5414.
\item[C. Kiefer] (1993b): ``Semiclassical gravity
with non-minimally coupled fields", in preparation.
\item[C. Kiefer] (1993c): ``Semiclassical gravity and the problem
 of time", to appear in the Proceedings of the Cornelius Lanczos International
   Centenary Conference.
\item[C. Kiefer] (1993d): ``Quantum cosmology and the emergence of
 a classical world", in {\em Mathematics, Philosophy and Modern
 Physics}, ed. by E. Rudolph and I.-O. Stamatescu (Springer, Berlin).
\item[C. Kiefer and T. P. Singh] (1991): Quantum Gravitational
 Corrections to the Functional Schr\"odinger Equation.
   Phys. Rev. D {\bf 44}, 1067.
\item[C. Kiefer, T. Padmanabhan, and T. P. Singh] (1991):
 A comparison between semiclassical gravity and semiclassical
 electrodynamics. Class. Quantum Grav. {\bf 8}, L 185.
\item[C. Kiefer, R. M\"uller, and T. P. Singh] (1993): Quantum Gravity
 and Non-unitarity in Black Hole Evaporation.
 Preprint gr-qc/9308024.
\item[J. Kowalski-Glikman and J. C. Vink] (1990): Gravity-matter
 mini-su\-per\-space: quantum regime, classical regime and in between.
 Class. Quantum Grav. {\bf 7}, 901.
\item[K. V. Kucha\v{r}] (1970): Ground State Functional of the
 Linearized Gravitational Field. Journ. Math. Phys. {\bf 11}, 3322.
\item[K. V. Kucha\v{r}] (1992): ``Time and Interpretations of Quantum Gravity",
 In {\em Proceedings of the 4th Canadian Conference on General Relativity
  and Relativistic Astrophysics}, ed. by G. Kunstatter, D. Vincent
   and J. Williams (World Scientific, Singapore, 1992).
\item[C. Kuo and L. H. Ford] (1993): Semiclassical gravity theory
and quantum fluctuations. Phys. Rev. D {\bf 47}, 4510.
\item[C. L\"ammerzahl] (1993): Private communication.
\item[L. D. Landau and E. M. Lifshitz] (1975): {\em Quantenmechanik}
 (Akademie-Ver\-lag, Berlin).
\item[V. G. Lapchinsky and V. A. Rubakov] (1979): Canonical
 quantization of gravity and quantum field theory in curved
 space-time. Acta Physica Polonica {\bf B}10, 1041.
\item[J. Louko] (1993): Holomorphic quantum mechanics with a
  quadratic Hamiltonian constraint. Phys. Rev. D {\bf 48}, 2708.
\item[V. Moncrief and C. Teitelboim] (1972): Momentum Constraints
 as Integrability Conditions for the Hamiltonian Constraint
 in General Relativity. Phys. Rev. D {\bf 6}, 966.
\item[D. N. Page] (1993): ``Black Hole Information", to appear in
 the {\em Proceedings of the 5th Canadian Conference on General
 Relativity and Relativistic Astrophysics}, Waterloo, Ontario,
 May 1993.
\item[T. Padmanabhan] (1990): A Definition for Time in Quantum
 Cosmology. Pramana {\bf 35}, L199.
\item[T. Padmanabhan and T. P. Singh]: On the semiclassical limit
  of the Wheeler-DeWitt equation. Class. Quantum Grav. {\bf 7},
   411 (1990).
\item[J. P. Paz and S. Sinha] (1991): Decoherence and back reaction:
      The origin of the semiclassical Einstein equations.
      Phys. Rev. D {\bf 44}, 1038.
\item[A. Peres] (1962): On Cauchy's Problem in General Relativity.
   Nuovo Cimento {\bf XXVI}, 53.
\item[E. Schr\"odinger] (1926a): Quantisierung als Eigenwertproblem
  (Erste Mitteilung). Annalen der Physik {\bf 79}, 361.
\item[E. Schr\"odinger] (1926b): Quantisierung als Eigenwertproblem
  (Vierte Mitteilung). Annalen der Physik {\bf 81}, 109.
\item[T. P. Singh] (1990): Gravity induced corrections to
 quantum mechanical wave functions. Class. Quantum Grav. {\bf 7},
 L149.
\item[T. P. Singh] (1993): ``Semiclassical gravity", in
  {\em Advances in Gravitation and Cosmology}, ed. by
  B. R. Iyer, A. R. Prasanna, R. K. Varma, and
  C. V. Vishveshwara (Wiley, Eastern, New Delhi).
\item[E. J. Squires] (1991): The dynamical role of time in quantum cosmology.
   Phys. Lett. {\bf A155}, 357.
\item[A. Vilenkin] (1989): Interpretation of the wave function
  of the Universe. Phys. Rev. D {\bf 39}, 1116.
\item[J. Wudka] (1990): Remarks on the Born-Oppenheimer
   approximation. Phys. Rev. D {\bf 41}, 712 (1990).
\item[H. D. Zeh] (1967): Symmetrieverletzende Modellzust\"ande
 und kollektive Bewegungen. Z. Phys. {\bf 202}, 38.
\item[H. D. Zeh] (1986): Emergence of Classical Time from a
             Universal Wave Function. Phys. Lett. {\bf A116}, 9.
\item[H. D. Zeh] (1988): Time in Quantum Gravity.
                  Phys. Lett. {\bf A126}, 311.
\item[H. D. Zeh] (1992): {\em The Physical Basis of The Direction
                    of Time} (Springer, Berlin).
\item[H. D. Zeh] (1993a): There are no quantum jumps, nor are
   there particles! Phys. Lett. {\bf A172}, 189.
\item[H. D. Zeh] (1993b): ``Decoherence and measurements",
  to appear in the Proceedings of {\em Stochastic evolution
  of quantum states in open systems and measurement processes},
  Budapest, March 1993.
\item[W. H. Zurek] (1991): Decoherence and the Transition from
 Quantum to Classical. Phys. Today {\bf 44}, 36.
\item[W. H. Zurek] (1993): Preferred States, Predictability,
  Classicality and the En\-viron\-ment-Induced Decoherence.
  Progr. Theor. Phys. {\bf 89}, 281.

\end{description}

\end{document}